\documentclass{IEEEtran}
\usepackage{amsfonts}
\usepackage{mathrsfs}
\usepackage{amssymb,amsmath}

\usepackage{algorithm}
\usepackage{algorithmic}
\usepackage{amsmath,amssymb,epsfig,graphics,subfigure}
\usepackage{theorem}
\usepackage{array,color}
\usepackage[compress]{cite}

\usepackage{graphicx}

\usepackage{flushend}

\newcounter{MYtempeqncnt}

\theoremheaderfont{\normalfont\bfseries}

\begin{document}

\title{Robust Transceiver with Tomlinson-Harashima Precoding for
Amplify-and-Forward MIMO Relaying Systems}

\author{Chengwen Xing, Minghua Xia, Feifei Gao, and Yik-Chung Wu

\thanks{Manuscript received August 15, 2011; revised March 6, 2012; Accepted May 4, 2012.  This work was supported in part by National Natural Science Foundation of China under
Grant No. 61101130, Sino-Swedish IMT-Advanced and Beyond Cooperative Program under Grant
No.2008DFA11780, Grant GRF HKU 7191/11E, the Specialized Research Fund for the Doctoral Program of
Higher Education of China (No. 20110002120059) and by the open research fund of National Mobile
Communications Research Laboratory, Southeast University (No. 2011D02).}
\thanks{C. Xing is with the School of Information and Electronics, Beijing Institute of Technology, Beijing, China (e-mail: chengwenxing@ieee.org). }
\thanks{
M. Xia is with the Division of Physical Sciences and Engineering, King
Abdullah University of Science and Technology, Thuwal, Saudi
Arabia (e-mail: minghua.xia@ieee.org). }
\thanks{
F. Gao is with Tsinghua National Laboratory for Information Science and Technology, Beijing, China and
is with National Mobile Communications Research Laboratory, Southeast University, Nanjing, China, and
is also with the School of Engineering and Science, Jacobs University, Bremen, Germany (e-mail:
feifeigao@ieee.org). }

\thanks{
Y.-C. Wu is with the Department of Electrical and Electronic Engineering, The University of Hong Kong, Hong Kong (e-mail:  ycwu@eee.hku.hk). }

}

\maketitle

\begin{figure*}[!t]
\centering
\includegraphics[width=0.7\textwidth]{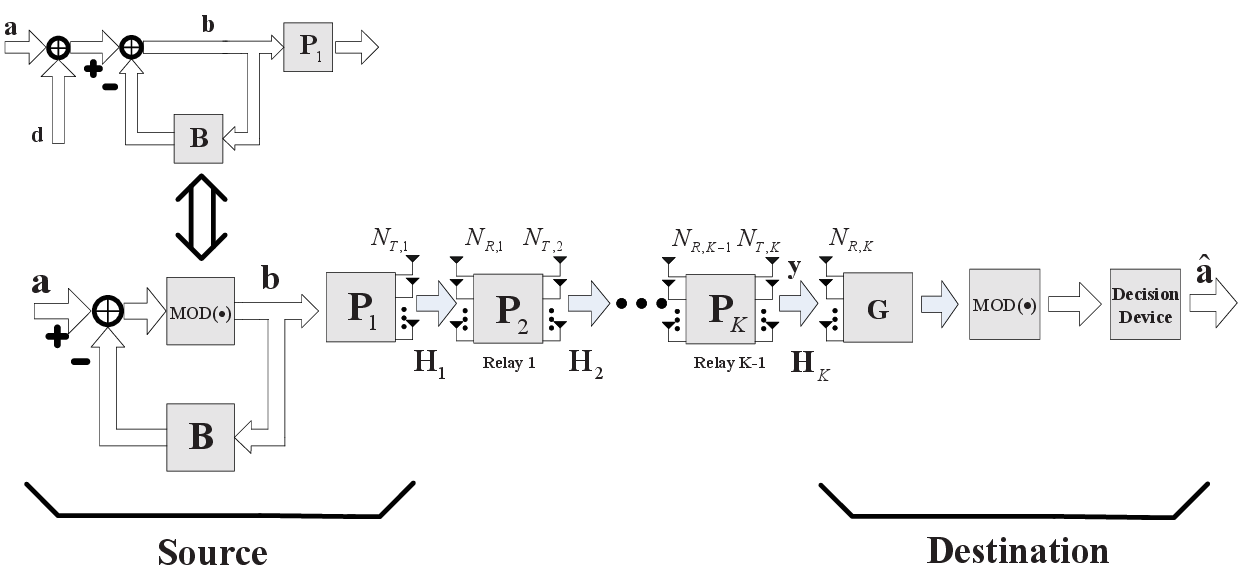}
\caption{Multi-hop AF MIMO relaying system with Tomlinson-Harashima Precoding at the
source.}\label{fig:1}
\end{figure*}

\begin{abstract}

In this paper, robust transceiver design with Tomlinson-Harashima precoding (THP) for multi-hop
amplify-and-forward (AF) multiple-input multiple-output (MIMO) relaying systems is investigated. At
source node, THP is adopted to mitigate the spatial intersymbol interference. However, due to its
nonlinear nature, THP is very sensitive to channel estimation errors. In order to reduce the effects
of channel estimation errors, a joint Bayesian robust design of THP at source, linear forwarding
matrices at relays and linear equalizer at destination is proposed. With novel applications of
elegant characteristics of multiplicative convexity and matrix-monotone functions, the optimal
structure of the nonlinear transceiver is first derived. Based on the derived structure, the
transceiver design problem reduces to a much simpler one with only scalar variables which can be
efficiently solved. Finally, the performance advantage of the proposed robust design over non-robust
design is demonstrated by simulation results.

\end{abstract}


\section{Introduction}
\label{sect:intro}

\IEEEPARstart{T}ransceiver design for amplify-and-forward (AF) multiple-input multiple-output (MIMO) relaying systems attracted a lot of attention recently, as it has a great potential to enhance the communication range of a simple point-to-point system, while providing spatial diversity and multiplexing gains.
AF MIMO relaying systems have a broad range of potential applications including resource exploration, vehicle communications, military ad hoc networks, satellite communications, etc \cite{Swami2007}. This system has also been considered to be adopted in the emerging wireless systems, such as LTE-Advanced and WINNER project.

Linear transceiver design for dual-hop AF MIMO relaying systems has been extensively studied in \cite{Medina07,Tang07,Guan08,Mo09,Rong09,Tseng09,
Xing10,Xing1012,XingGlobecom,XingICSPCC,Rong2009TWC}. In particular, joint design of relay forwarding matrix and destination equalizer minimizing mean-square-error (MSE) of data streams is discussed in \cite{Guan08}. Joint design of source precoder, relay forwarding matrix and destination equalizer minimizing MSE is investigated in \cite{Mo09,Tseng09,Rong09}. The capacity maximization transceiver design has also been reported in \cite{Tang07,Rong09,Medina07}. On the other hand, linear transceiver design for multi-hop AF MIMO relaying systems with prefect channel state information (CSI) is discussed in \cite{Rong2009TWC}. Furthermore, robust design, which takes channel estimation errors into account, is recently investigated in \cite{Xing10,Xing1012,XingGlobecom,XingICSPCC}, where the channel estimation uncertainty is considered as nuisance parameters and removed in Bayesian sense.

In general, there are two goals in transceiver designs: transmitting as much information as possible and recovering the signal at receiver as accurately as possible. The latter one is the starting point of this paper. For multiple-antenna systems with fixed bit rates, it is well-known that nonlinear transceivers usually have performance advantage in terms of bit error rate (BER) than their linear counterparts \cite{Amico2008,Jiang2005,Shenouda2008}. Recently, nonlinear transceiver design for AF MIMO relaying systems assuming perfect CSI, was introduced in \cite{Rong2011}. There are two kinds of nonlinear transceiver design: decision-feedback equalization (DFE) based design and Tomlinson-Harashima precoding (THP) based design. In fact, there exists a duality between these two designs \cite{Rong2011,Palomar2006}. However, as THP is performed at transmitter, it is free of error propagation compared to DFE based one. THP is the transmitter counterpart of the vertical BELL-Labs Layered Space-Time (V-BLAST) system. THP can effectively mitigate intersymbol interference or multi-user interference, and is also widely used as one-dimensional dirty paper coding (DPC). Due to its nonlinear nature, unfortunately, THP is more sensitive to channel estimation errors than its linear counterpart. In the presence of channel estimation errors, the performance of THP would degrade severely \cite{Dietrich2007}. Therefore, robust nonlinear transceiver design is a promising way to mitigate such problem.  This is the motivation of the current work.

In this paper, we consider a general multi-hop AF MIMO relaying system. The THP at the source, linear forwarding matrices at multiple relays and linear destination equalizer matrix are jointly optimized under channel estimation errors at all terminals. As in this case many design objectives of THP can be considered as a multiplicatively Schur-convex or multiplicatively Schur-concave function, in this work, a unified optimization problem is investigated whose objective functions are multiplicative Schur-convex/concave. With novel applications of results in multiplicative Schur-convexity and  matrix-monotone functions, the optimal diagonal structure of the transceiver is derived. With the obtained optimal structures, the transceiver design is then significantly simplified and then iterative water-filling alike solutions are adopted to solve for the remaining unknown variables. It is found that if the objective function is multiplicatively Schur-concave, the proposed nonlinear transceiver design reduces to linear transceiver design. The performance advantage of the proposed robust design is assessed by simulations and is shown to perform much better than the corresponding non-robust design. Notice that while delay is a critical consideration for relaying communication, in this paper, we assume that the network size is limited and the effects of time delay in transmission are not considered.

The following notations are used throughout this paper. Boldface
lowercase letters denote vectors, while boldface uppercase letters
denote matrices. The notation ${\bf{Z}}^{\rm{H}}$ denotes the
Hermitian of the matrix ${\bf{Z}}$, and ${\rm{Tr}}({\bf{Z}})$ is the
trace of the matrix ${\bf{Z}}$. The symbol ${\bf{I}}_{N}$ denotes an
$N \times N$ identity matrix. The notation ${\bf{Z}}^{1/2}$ is the
Hermitian square root of the positive semidefinite matrix
${\bf{Z}}$, such that ${\bf{Z}}^{1/2}{\bf{Z}}^{1/2}={\bf{Z}}$ and
${\bf{Z}}^{1/2}$ is also a Hermitian matrix.
The symbol ${\mathbb{E}}\{\bullet\}$ represents the statistical expectation. For two Hermitian matrices, ${\bf{C}} \succeq
{\bf{D}}$ means that ${\bf{C}}-{\bf{D}}$ is a positive semi-definite
matrix. The $(n,m)^{\rm{th}}$ entry of a matrix ${\bf{Z}}$ is denoted as
$[{\bf{Z}}]_{n,m}$ and ${\boldsymbol \lambda}({\bf{Z}})$ represents the vector consisting of the eigenvalues of ${\bf{Z}}$.

\section{Signal Model and Problem Formulation}
\label{Section:signal_model}
\subsection{Signal Model}
In this paper, a $K$-hop amplify-and-forward MIMO relaying system is investigated, in which there is one source, one destination and $K-1$ relays, as shown in Fig.~\ref{fig:1}. The source is equipped with $N_{T,1}$ transmit antennas. The $k^{\rm{th}}$ relay has
$N_{R,k}$ receive antennas and $N_{T,k+1}$ transmit antennas. The destination is equipped with $N_{R,K}$ receive antennas. At the source, at each time slot, there is a $N\times1$ vector ${\bf{a}}=[a_1,a_2,\cdots,a_N]^{\rm{T}}$ to be transmitted. Specifically, the data symbols are chosen from M-QAM constellation with the real and imaginary parts of $a_k$ belong to the set ${\mathcal{A}}=\{ \pm 1,\pm3,\cdots, \pm (\sqrt{\mathcal{M}}-1)\}$ \footnote{In this paper, only square QAM is considered.}.

As shown by Fig.~\ref{fig:1}, at the transmitter, the data vector ${\bf{a}}$ is fed into the a
precoding unit which consists of a $N\times N$ feedback matrix ${\bf{B}}$ and a nonlinear modulo
operator ${\rm{MOD}}_{\mathcal{M}}(\bullet)$. The square matrix ${\bf{B}}$ is a strictly lower
triangular matrix which allows data precoding in a recursive fashion and the
${\rm{MOD}}_{\mathcal{M}}(\bullet)$ is defined as
\begin{align}
&{\rm{MOD}}_{\mathcal{M}}(x) \nonumber\\ &=x
-2\sqrt{{\mathcal{M}}}\left[\left\lfloor\frac{{\rm{Re}}(x)}{2\sqrt{{\mathcal{M}}}}+\frac{1}{2}\right\rfloor
+\sqrt{-1}\left\lfloor\frac{{\rm{Im}}(x)}{2\sqrt{{\mathcal{M}}}}+\frac{1}{2} \right\rfloor\right],
\end{align}where the symbol $\lfloor z\rfloor$ denotes the
largest integer not exceeding $z$. The nonlinear modulo operator
 reduces the output signals into a square region
 $[-\sqrt{{\mathcal{M}}},\sqrt{{\mathcal{M}}})\times [-\sqrt{{\mathcal{M}}},\sqrt{{\mathcal{M}}})$.
In the equation, ${\rm{Re}}(x)$ and ${\rm{Im}}(x)$ denote the real and imaginary parts of $x$, respectively.

%
%
%
%

Generally speaking, nonlinear operation is more complicated to be analyzed than linear operation. To simplify the following analysis, as shown by Fig.~\ref{fig:1}, the nonlinear precoder can be interpreted as the following linear operation as
\begin{align}
\label{b_THP}
b_k=a_k-\sum_{l=1}^{k-1}[{\bf{B}}]_{k,l}b_l+d_k
\end{align}where $d_k=2\sqrt{{\mathcal{M}}}I_k$ and $I_k$ is a complex number whose real and imaginary components are both integer. While we do not need to know the exact value of $d_k$, it has the effect of reducing $b_k$ into the square region $[-\sqrt{{\mathcal{M}}},\sqrt{{\mathcal{M}}})\times [-\sqrt{{\mathcal{M}}},\sqrt{{\mathcal{M}}})$. The previous equation can be written into a compact form as\begin{align}
\label{C}
{\bf{b}}=(\underbrace{{\bf{B}}+{\bf{I}}_N}_{\triangleq {\bf{C}}})^{-1}(\underbrace{{\bf{a}}+{\bf{d}}}_{\triangleq {\bf{s}}})
\end{align}where ${\bf{b}}\triangleq[b_1,\cdots,b_N]^{\rm{T}}$, ${\bf{d}}\triangleq[d_1,\cdots,d_N]^{\rm{T}}$, and ${\bf{C}}$ is a lower triangular matrix with unit diagonal elements, i.e., $[{\bf{C}}]_{k,l}=0$ for $k < l$ and $[{\bf{C}}]_{k,k}=1$.

 After the nonlinear operation, the vector ${\bf{b}}$ is multiplied with a precoder matrix ${\bf{P}}_1$ under a transmit power constraint ${\rm{Tr}}({\bf{P}}_1{\bf{R}}_{\bf{b}}{\bf{P}}_1^{\rm{H}})\le P_1$ where $P_1$ is the maximum transmit power at the source. When the elements of ${\bf{a}}$ are independent and identically distributed (i.i.d.) over the constellation and the dimension of modulation constellation  $\mathcal{M}$ is large, ${\bf{b}}$ can be considered as i.i.d. \cite{Fischer2002}, i.e.,
 \begin{align}
 \label{R_b_THP}
 {\bf{R}}_{{\bf{b}}} ={2({\mathcal{M}}-1)}/{3} {\bf{I}}_N\triangleq\sigma_b^2{\bf{I}}_N.
\end{align}

The received signal ${\bf{x}}_1$ at the first relay is formulated as
\begin{align}
{\bf{x}}_1={\bf{H}}_{1}{\bf{P}}_1{\bf{b}}+{\bf{n}}_1
\end{align}where ${\bf{H}}_1$ is the channel between the source and the first relay and ${\bf{n}}_1$ is additive Gaussian noise with mean zero and covariance matrix ${\bf{R}}_{{\bf{n}}_1}=\sigma_{n_1}^2{\bf{I}}_{N_{R,1}}$.

At the first relay, the received signal ${\bf{x}}_1$ is multiplied by a forwarding matrix ${\bf{P}}_2$ and then the resultant signal is
transmitted to the second relay. The received signal at
the second relay can be written as
\begin{align}
{\bf{x}}_2={\bf{H}}_{2}{\bf{P}}_{2}{\bf{H}}_{1}{\bf{P}}_1{\bf{b}}
+{\bf{H}}_{2}{\bf{P}}_{2}{\bf{n}}_1+{\bf{n}}_2
\end{align}where ${\bf{H}}_{2}$ is the MIMO channel matrix between the first and second relay, and ${\bf{n}}_2$ is the additive Gaussian noise
vector at the second hop with zero mean and covariance matrix ${\bf{R}}_{{\bf{n}}_2}=\sigma_{n_2}^2{\bf{I}}_{N_{R,2}}$.
Similarly, at the $k^{\rm{th}}$ relay the received signal is
\begin{align}
{\bf{x}}_k={\bf{H}}_k{\bf{P}}_k{\bf{x}}_{k-1}+{\bf{n}}_{k}
\end{align} with ${\bf{H}}_k$ and ${\bf{n}}_k$ are the channel and additive noise at the $k^{\rm{th}}$  hop, respectively. In this paper, we considered slow fading channels with ${\bf{H}}_k$ being fixed in each transmission.

The covariance matrix of ${\bf{n}}_k$ is denoted as ${\bf{R}}_{{\bf{n}}_k}=\sigma_{n_k}^2{\bf{I}}_{N_{R,k}}$. Finally, for a $K$-hop AF MIMO relaying system, the received signal at the destination is
\begin{align}
{\bf{y}} = \left[{\prod_{k=1}^K}({\bf{H}}_{k}{\bf{P}}_k)\right]{\bf{b}}  + \sum_{k=1}^{K-1}\left\{ \left[\prod_{l={k+1}}^K({\bf{H}}_{l}{\bf{P}}_l)\right]{\bf{n}}_k
\right\}+{\bf{n}}_K,
\end{align}where ${\prod_{k=1}^K}{\bf{Z}}_k$ denotes ${\bf{Z}}_K\times \cdots \times {\bf{Z}}_1$. In order to guarantee the transmitted data ${\bf{s}}$ can be
recovered at the destination, it is assumed that
$N_{T,k}$ and $N_{R,k}$ are greater than or equal to $N$ \cite{Guan08}.

In practice, the channels ${\bf{H}}_k$ are estimated and channel estimation errors are inevitable. Therefore, the channel ${\bf{H}}_k$ can be expressed as
\begin{align}
\label{channel_state}
{\bf{H}}_k={\bf{\bar H}}_k+\Delta{\bf{H}}_k,
\end{align}where ${\bf{\bar H}}_{k}$ is the estimated channels, and $\Delta{\bf{H}}_{k}$ is the corresponding channel estimation errors\footnote{In this paper, only channel estimation errors are taken into account.} whose elements are zero mean Gaussian random variables. Furthermore, the $N_{R,k} \times
N_{T,k}$ matrix $\Delta{\bf{H}}_{k}$  can be decomposed using the widely used Kronecker model \cite{Zhang08,Xing10,Xing1012} as
$\Delta{\bf{H}}_{k}={\boldsymbol{\Sigma}}
_{k}^{{1}/{2}}{\bf{H}}_{W,k}{\boldsymbol{\Psi}}_{k}^{{1}/{2}}$, where the elements of the $N_{R,k} \times N_{T,k}$ matrix
${\bf{H}}_{W,k}$ are i.i.d. Gaussian random variables with zero mean and unit variance. The specific formulas of ${\boldsymbol \Sigma}_k$ and ${\boldsymbol \Psi}_k$ are determined by the training sequences and channel estimators \cite{Ding09,Xing1012,XingGlobecom,Xing10}.

\subsection{Problem Formulation}
As shown by Fig.~\ref{fig:1},  at the destination, a linear equalizer ${\bf{G}}$  is adopted and is followed by a modulo operator. As the real and imaginary parts of ${\bf{d}}$ are both integer multiples of $2\sqrt{{\mathcal{M}}}$, the effect of ${\bf{d}}$ will be perfectly removed by modulo operator at the destination. As a result, estimating ${\bf{s}}$ is equivalent to estimating ${\bf{a}}$ \cite{Amico2008,Shenouda2008}. Thus at the destination, a linear equalizer ${\bf{G}}$ is used to detect the data vector ${\bf{s}}$. The MSE
matrix of the data vector is defined as $\mathbb{E}\{({\bf{G}}{\bf{y}}-{\bf{s}})({\bf{G}}{\bf{y}}-{\bf{s}})^{\rm{H}}\}$ \cite{Amico2008,Fischer2002}, where the expectation is taken with respect to random data, channel estimation errors, and noise. Following a similar derivation to that in \cite{Xing1012}, it can be shown that
\begin{align}
&\label{MSE_0} {\boldsymbol{\Phi}}({\bf{G}},\{{\bf{P}}_k\}_{k=1}^K,{\bf{C}}) \nonumber \\
&={\mathbb{E}}\{({\bf{G}}{\bf{y}}-{\bf{C}}{\bf{b}})({\bf{G}}{\bf{y}}-{\bf{C}}{\bf{b}})^{\rm{H}}\} \nonumber \\
&={\bf{G}}[{\bf{\bar H}}_{K}{\bf{P}}_K{\bf{R}}_{{\bf{x}}_{K-1}}{\bf{P}}_K^{\rm{H}}{\bf{\bar
H}}_{K}^{\rm{H}} +{\rm{Tr}}({\bf{P}}_K{\bf{R}}_{{\bf{x}}_{K-1}}{\bf{P}}_K^{\rm{H}}{\boldsymbol
\Psi}_K){\boldsymbol \Sigma}_K \nonumber\\
&  \quad +{\bf{R}}_{n_K}]{\bf{G}}^{\rm{H}} -\sigma_b^2{\bf{G}}\prod_{k=1}^K\left({\bf{\bar
H}}_k{\bf{P}}_k\right){\bf{C}}^{\rm{H}} \nonumber\\
&\ \ \ \  \ \ \ \ \ \ \ \ \quad -\sigma_b^2\left[{\bf{G}}\prod_{k=1}^K\left({\bf{\bar
H}}_k{\bf{P}}_k\right){\bf{C}}^{\rm{H}}\right]^{\rm{H}}+\sigma_b^2{\bf{C}}{\bf{C}}^{\rm{H}}
\end{align}where matrices ${\bf{R}}_{{\bf{x}}_k}$ is defined as\begin{align}
\label{R_x} {\bf{R}}_{{\bf{x}}_k}&\triangleq \mathbb{E}\{{\bf{x}}_k{\bf{x}}_k^{\rm{H}}\}
\nonumber\\
&={\bf{\bar H}}_{k}{\bf{P}}_k{\bf{R}}_{{\bf{x}}_{k-1}}{\bf{P}}_k^{\rm{H}}{\bf{\bar H}}_{k}^{\rm{H}}
\!+\! {\rm{Tr}}({\bf{P}}_k{\bf{R}}_{{\bf{x}}_{k-1}}{\bf{P}}_k^{\rm{H}}{\boldsymbol
\Psi}_{k}){\boldsymbol \Sigma}_{k}\!+\!\!{\bf{R}}_{{\bf{n}}_k}.
\end{align}It is obvious that ${\bf{R}}_{{\bf{x}}_k}$ is the covariance matrix of the received signal at the relay. Notice that ${\bf{R}}_{{\bf{x}}_0}={\bf{R}}_{\bf{b}}=\sigma_b^2{\bf{I}}_N$.

 For MIMO transceiver design, a wide range of objective functions can be expressed as a function of the diagonal elements of the MSE matrix. For example, for sum MSE minimization, the objective function is ${{f}}([{\rm{MSE}}_1,\cdots,{\rm{MSE}}_N]^{\rm{T}})=\sum_{n=1}^N{\rm{MSE}}_n$, where ${\rm{MSE}}_n= [{\boldsymbol{\Phi}}({\bf{G}},\{{\bf{P}}_k\}_{k=1}^K,{\bf{C}})]_{n,n}$. For product MSE minimization, the objective function is ${{f}}([{\rm{MSE}}_1,\cdots,{\rm{MSE}}_N]^{\rm{T}})=\prod_{n=1}^N{\rm{MSE}}_n.$ Furthermore, worst-case MSE minimization corresponds to minimizing the objective function given as $
{{f}}([{\rm{MSE}}_1,\cdots,{\rm{MSE}}_N]^{\rm{T}})=\max_{n=1,2,\cdots,N}\{ {\rm{MSE}}_n \}$ \cite{Amico2008,Palomar03,Rong09,Jiang2005}. On the other hand, weighted geometric mean MSE minimization corresponds to minimizing the following objective function ${{f}}([{\rm{MSE}}_1,\cdots,{\rm{MSE}}_N]^{\rm{T}})=\prod_{n=1}^N{\rm{MSE}}_n^{w_n}$ with $ w_1\ge w_2\cdots\ge w_N \ge 0$. Therefore, a unified transceiver design optimization problem can be formulated as
\begin{align}
\label{Problem_0}
& \min_{{\bf{G}},{\bf{P}}_k,{\bf{C}}} \ \ \ {{f}}([{\rm{MSE}}_1,\cdots,{\rm{MSE}}_N]^{\rm{T}}) \nonumber \\
& \ \ {\rm{s.t.}} \ \ \ \ \ \ {\rm{MSE}}_n= [{\boldsymbol{\Phi}}({\bf{G}},\{{\bf{P}}_k\}_{k=1}^K,{\bf{C}})]_{n,n} \nonumber \\
& \ \ \ \ \ \ \ \ \ \ \ \ {\rm{Tr}}({\bf{P}}_{k}{\bf{R}}_{{\bf{x}}_{k-1}}{\bf{P}}_k^{\rm{H}})\le P_k, \ \ k=1,\cdots,K
\end{align}where the matrix ${\bf{C}}$ is a lower triangular matrix with unit diagonal elements and $P_k$ is the maximum transmit power at the $k^{\rm{th}}$ node.

In general, the objective function ${f}(\bullet)$ possesses two important properties:

\noindent (1) ${f}(\bullet)$ is an increasing real-valued vector function $\mathbb{C}^{N}\rightarrow \mathbb{R} $, i.e., for two vectors ${\bf{u}}=[u_1,u_2,\cdots,u_N]^{\rm{T}}$ and ${\bf{v}}=[v_1,v_2,\cdots,v_N]^{\rm{T}}$, when ${{u}}_n\ge {{v}}_n$, we have ${f}({\bf{u}})\ge f({\bf{v}})$.
This property is natural in transceiver design.  This is because for two designs resulting in $[{\rm {MSE}}_1,\cdots, {\rm{MSE}}_N]^{\rm{T}}$ and $[{\rm {\widetilde{ MSE}}}_1,\cdots, {\rm{\widetilde{ MSE}}}_N]^{\rm{T}}$, suppose ${\rm{MSE}}_n<{\rm{\widetilde {MSE}}}_n$ for all $n$, we will prefer the former design. This fact is reflected in ${f}(\bullet)$ being an increasing function.

\noindent (2) ${f}(\bullet)$ is multiplicatively Schur-convex or concave, with definitions given below.

%




\noindent \textbf{Definition 1:} For any ${\bf{z}} \in {\mathbb{R}}^{n}$, let ${{z}}_{[k]}$ denotes the $k^{\rm{th}}$ largest elements of ${\bf{z}}$ and ${{z}}_{(k)}$ denotes the $k^{\rm{th}}$ smallest elements of ${\bf{z}}$, i.e.,
${{z}}_{[1]}\ge \cdots \ge {{z}}_{[N]} \ \text{and} \ {{z}}_{(1)}\le \cdots \le {{z}}_{(N)}$. For two vectors ${\bf{v}},{\bf{u}}$ whose elements are \textbf{nonnegative}, ${\bf{v}}\prec_{\times} {\bf{u}}$ is defined as
\begin{align}
\label{def_2} &\prod_{i=1}^k {{v}}_{[i]}\le \prod_{i=1}^k {{u}}_{[i]}, \  k=1,\cdots,N-1 \ \text{and}
\ \prod_{i=1}^N {{v}}_{[i]}= \prod_{i=1}^N {{u}}_{[i]}.
\end{align}

\noindent \textbf{Definition 2:} A function ${\phi}(\bullet)$ is  multiplicatively Schur-convex if and only if ${\bf{v}}\prec_{\times} {\bf{u}}$ implies ${\phi}({\bf{v}})\le {\phi}({\bf{u}})$. Notice that ${\phi}(\bullet)$ is multiplicatively Schur-convex if and only if $-{\phi}(\bullet)$ is multiplicatively  Schur-concave.

Notice that \textbf{Definition 2} cannot be directly used to prove whether a function is  multiplicatively Schur-convex or Schur-concave. In practice, we need the following \textbf{Lemma 1}.

\noindent \textbf{Lemma 1:} Let $\phi(\bullet)$ be a continuous real-valued function defined on $\mathcal{D}= \{{\bf{z}}: z_1\ge \cdots \ge z_N \ge 0\}$.
 Then ${\phi}(\bullet)$ is multiplicatively  Schur-convex if and only if
  for all ${\bf{z}} \in \mathcal{D}$,
\[{\phi}(z_1,\cdots,z_{k-1},z_{k}/e, z_{k+1}\times e,z_{k+2},\cdots,z_N)
\]is decreasing in $e$ over the following regions
\begin{align}
& 1\le e \ \text{and} \  z_{k}/e  \ge  z_{k+1}\times e   \ \ \text{for} \ \ k=1,\cdots,N-1.
\end{align}

\noindent \textsl{\textbf{Proof:}} See Appendix~\ref{Appedix:1}. $\blacksquare$

 With \textbf{Lemma 1} and straightforward computation, it can be proved that the four objective functions mentioned above are multiplicatively Schur-convex or concave. In the following, for notational convenience, multiplicatively Schur-convex/concave is referred to as M-Schur-convex/concave.

\noindent \textsl{\textbf{Remark 1:}} Notice that in\cite{Shenouda2008,Amico2008}, there is another way to prove whether a function is M-Schur-convex/concave.  However, the method in \cite{Shenouda2008,Amico2008} requires all input variables $z_1, z_2,...,z_N >0$.  In contrast, {\textbf{Lemma 1}} provides a stronger result and allows elements of ${\bf{z}}$ being zero.

\noindent \textsl{\textbf{Remark 2:}} The differences between our work and \cite{Amico2008,Shenouda2008} are twofold. (a) The system considered in \cite{Amico2008,Shenouda2008} is a point-to-point MIMO system, while our work focuses on a multi-hop AF MIMO relaying system. (b) In the above two works, the involved CSI is perfectly known. In this paper, we consider a robust transceiver design under Gaussian distributed channel estimation errors.
Generally speaking, the problem tackled in this paper is more complicated and more challenging, because of more variables, more constraints, a more complicated objective function.

\section{Optimal Design of ${\bf{G}}$ and ${\bf{C}}$}
The linear minimum mean-square-error (LMMSE) equalizer is obtained by setting the differentiation of the trace of (\ref{MSE_0}) with respect to ${\bf{G}}^*$ (the conjugate of ${\bf{G}}$) to be zero, and we have
\begin{align}
\label{G_opt}
{\bf{G}}_{\rm{LMMSE}}&=\sigma_b^2\left[\prod_{k=1}^K\left({\bf{\bar H}}_k{\bf{P}}_k\right){\bf{C}}^{\rm{H}}\right]^{\rm{H}}\!\![{\bf{\bar H}}_{K}{\bf{P}}_K{\bf{R}}_{{\bf{x}}_{K-1}}{\bf{P}}_K^{\rm{H}}{\bf{\bar H}}_{K}^{\rm{H}}\nonumber \\
&\ \ \ \ +{\rm{Tr}}({\bf{P}}_K{\bf{R}}_{{\bf{x}}_{K-1}}{\bf{P}}_K^{\rm{H}}{\boldsymbol
\Psi}_K){\boldsymbol \Sigma}_K+{\bf{R}}_{{\bf{n}}_K}]^{-1}.
\end{align}In terms of MSE, LMMSE estimator is a dominated estimator in linear estimators \cite{Palomar03}, i.e.,
\begin{align}
{\boldsymbol{\Phi}}({\bf{G}}_{\rm{LMMSE}},\{{\bf{P}}_k\}_{k=1}^K,{\bf{C}})
\preceq{\boldsymbol{\Phi}}({\bf{G}},\{{\bf{P}}_k\}_{k=1}^K,{\bf{C}})
\end{align}which implies
\[
[{\boldsymbol{\Phi}}({\bf{G}}_{\rm{LMMSE}},\{{\bf{P}}_k\}_{k=1}^K,{\bf{C}})]_{n,n}
\le
[{\boldsymbol{\Phi}}({\bf{G}},\{{\bf{P}}_k\}_{k=1}^K,{\bf{C}})]_{n,n}.\] As $f(\bullet)$ is an increasing function, and there is no constraint on ${\bf{G}}$ in (\ref{Problem_0}), the optimal linear equalizer is LMMSE equalizer, i.e., ${\bf{G}}_{\rm{opt}}={\bf{G}}_{\rm{LMMSE}}$.

Substituting the optimal equalizer (\ref{G_opt}) into the MSE formulation (\ref{MSE_0}), the MSE matrix is rewritten as
\begin{align}
&{\boldsymbol{\Phi}}_{\rm{MSE}}(\{{\bf{P}}_k\}_{k=1}^K,{\bf{C}})  \nonumber \\&
=\sigma_b^2{\bf{C}}\Bigg(\!{\bf{I}}_N \!- \!\sigma_b^2\left[\prod_{k=1}^K\left({\bf{\bar
H}}_k{\bf{P}}_k\right)\!\right]^{\rm{H}}\!\! [{\bf{\bar
H}}_{K}{\bf{P}}_K{\bf{R}}_{{\bf{x}}_{K-1}}  \nonumber
\\& \ \ \ \times {\bf{P}}_K^{\rm{H}}{\bf{\bar H}}_{K}^{\rm{H}}
+{\rm{Tr}}({\bf{P}}_K{\bf{R}}_{{\bf{x}}_{K-1}}{\bf{P}}_K^{\rm{H}}{\boldsymbol \Psi}_K){\boldsymbol
\Sigma}_K+{\bf{R}}_{n_K}]^{-1} \nonumber
\\& \quad\quad\quad\quad\quad\quad\quad\quad\quad\times \left[\prod_{k=1}^K\left({\bf{\bar
H}}_k{\bf{P}}_k\right)\right]\Bigg){\bf{C}}^{\rm{H}}
\end{align}based on which the optimization problem (\ref{Problem_0}) becomes
\begin{align}
\label{Problem_1}
& \min_{{\bf{P}}_k,{\bf{C}}} \ \ \ {{f}}([{\rm{MSE}}_1,\cdots,{\rm{MSE}}_N]^{\rm{T}}) \nonumber \\
& \ \ {\rm{s.t.}} \ \ \ \ {\rm{MSE}}_n= [{\boldsymbol{\Phi}}_{\rm{MSE}}(\{{\bf{P}}_k\}_{k=1}^K,{\bf{C}})]_{n,n} \nonumber \\
& \ \ \ \ \ \ \ \ \ \ {\rm{Tr}}({\bf{P}}_{k}{\bf{R}}_{{\bf{x}}_{k-1}}{\bf{P}}_k^{\rm{H}})\le P_k.
\end{align}

 From the definition of ${\bf{R}}_{\bf{{x}}_k}$ in (\ref{R_x}), it is noticed that ${\bf{R}}_{{\bf{x}}_k}$ is a function of ${\bf{P}}_l$ with $l\le k$. In other words, the constraints in (\ref{Problem_1}) are coupled with each other. In order to simplify the analysis, we define the following new variables
\begin{align}
& {\bf{F}}_1={\bf{P}}_1{\bf{R}}_{{\bf{b}}}^{1/2}{\bf{Q}}_0^{\rm{H}} \label{F_k_1} \\
{\text{and}} \ \ &{\bf{F}}_k={\bf{P}}_k{\bf{K}}_{{\bf{F}}_{k-1}}^{1/2}
({\bf{K}}_{{\bf{F}}_{k-1}}^{-1/2}{\bf{\bar H}}_{k-1}{\bf{F}}_{k-1}{\bf{F}}_{k-1}^{\rm{H}}{\bf{\bar H}}_{k-1}^{\rm{H}}{\bf{K}}_{{\bf{F}}_{k-1}}^{-1/2}
 \nonumber \\ & \quad \quad \quad \quad \quad \quad \quad \quad \quad \quad \quad  +{\bf{I}}_{N_{R,k-1}})^{1/2}{\bf{Q}}_{k-1}^{\rm{H}} \label{F_k_2}
\end{align}where ${\bf{K}}_{{\bf{F}}_k}$ is defined as\footnote{Putting the definition of ${\bf{F}}_k$ into (\ref{R_x}) and comparing (\ref{R_x}) with (\ref{K_F}), the matrix ${\bf{K}}_{{\bf{F}}_k}$ can be interpreted as the equivalent noise covariance matrix at the $k^{\rm{th}}$ hop.}
\begin{align}
\label{K_F}
{\bf{K}}_{{\bf{F}}_k}\triangleq {\rm{Tr}}({\bf{F}}_k{\bf{F}}_k^{\rm{H}}{\boldsymbol \Psi}_k){\boldsymbol \Sigma}_k+\sigma_{n_k}^2{\bf{I}}_{N_{R,k}},
\end{align}and the matrix ${\bf{Q}}_k$ is an additional unknown unitary matrix. Based on the definition of ${\bf{F}}_k$  in (\ref{F_k_1}) and (\ref{F_k_2}), it is easy to show that ${\bf{F}}_{k}{\bf{F}}_k^{\rm{H}}={\bf{P}}_k{\bf{R}}_{{\bf{x}}_{k-1}}{\bf{P}}_k^{\rm{H}}$ and thus the power constraints becomes
\begin{align}
\label{constraints}
{\rm{Tr}}({\bf{P}}_k{\bf{R}}_{{\bf{x}}_{k-1}}{\bf{P}}_k^{\rm{H}})={\rm{Tr}}({\bf{F}}_{k}{\bf{F}}_k^{\rm{H}})\le P_k.
\end{align}Therefore, in terms of the new variables ${\bf{F}}_k$, the power constraints become independent of each other, which facilitates further manipulations.

Meanwhile, using (\ref{F_k_1}) and (\ref{F_k_2}), the MSE matrix is further reformulated as
(\ref{object_MSE}) on the top of the page.
\begin{figure*}[!t]
\normalsize \setcounter{MYtempeqncnt}{\value{equation}} \setcounter{equation}{22}
\begin{align}
\label{object_MSE} {\boldsymbol{\Phi}}_{\rm{MSE}}({\bf{Q}}_k,\{{\bf{F}}_k\}_{k=1}^K,{\bf{C}})
&={\bf{C}}\Bigg({\bf{I}}_N-{\bf{Q}}_0^{\rm{H}} \bigg
\{\prod_{k=1}^K[{\bf{Q}}_k({\bf{K}}_{{\bf{F}}_{k}}^{-1/2}{\bf{\bar
H}}_{k}{\bf{F}}_{k}{\bf{F}}_{k}^{\rm{H}}{\bf{\bar
H}}_{k}^{\rm{H}}{\bf{K}}_{{\bf{F}}_k}^{-1/2}+{\bf{I}}_{N_{R,k}})^{-1/2}
{\bf{K}}_{{\bf{F}}_k}^{-1/2}{\bf{\bar H}}_{k}{\bf{F}}_{k}] \bigg \}^{\rm{H}} \nonumber \\
& \quad \times \bigg\{\prod_{k=1}^K[{\bf{Q}}_k\underbrace{({\bf{K}}_{{\bf{F}}_k}^{-1/2}{\bf{\bar
H}}_{k}{\bf{F}}_{k}{\bf{F}}_{k}^{\rm{H}}{\bf{\bar H}}_{k}^{\rm{H}}{\bf{K}}_{{\bf{F}}_k}^{-1/2}
+{\bf{I}}_{N_{R,k}})^{-1/2}{\bf{K}}_{{\bf{F}}_k}^{-1/2}{\bf{\bar H}}_{k}{\bf{F}}_{k}}_{\triangleq
{\bf{M}}_k}]\bigg\}{\bf{Q}}_0\Bigg){\bf{C}}^{\rm{H}}\sigma_b^2 \nonumber
\\&=\sigma_b^2{\bf{C}}({\bf{I}}_N- {\bf{Q}}_0^{\rm{H}}\underbrace{{\bf{M}}_1^{\rm{H}}
{\bf{Q}}_1^{\rm{H}}{\bf{M}}_2^{\rm{H}}{\bf{Q}}_{2}^{\rm{H}}\cdots{\bf{M}}_K^{\rm{H}}
{\bf{Q}}_{K}^{\rm{H}}{\bf{Q}}_{K}{\bf{M}}_K\cdots{\bf{Q}}_{2}{\bf{M}}_2{\bf{Q}}_{1}{\bf{M}}_1}_{\triangleq
{\boldsymbol \Theta}}{\bf{Q}}_0){\bf{C}}^{\rm{H}}.
\end{align}
\hrulefill \stepcounter{MYtempeqncnt} \setcounter{equation}{\value{MYtempeqncnt}} \vspace*{3pt}
\end{figure*}
Based on (\ref{constraints}) and (\ref{object_MSE}), the optimization problem (\ref{Problem_1}) is simplified as
\begin{align}
\label{OPT_F_C}
& \min_{{\bf{F}}_k,{\bf{Q}}_k,{\bf{C}}} \ \  {{f}}([{\rm{MSE}}_1,\cdots,{\rm{MSE}}_N]^{\rm{T}}) \nonumber \\
& \ \ \ {\rm{s.t.}} \ \ \ \ \ {\rm{MSE}}_n=\sigma_b^2\left[{\bf{C}}({\bf{I}}_N-{\bf{Q}}_0^{\rm{H}}{\boldsymbol \Theta}{\bf{Q}}_0){\bf{C}}^{\rm{H}}\right]_{n,n} \nonumber \\
& \ \ \ \ \ \ \ \ \ \ \ \ {\boldsymbol \Theta}={\bf{M}}_1^{\rm{H}}{\bf{Q}}_1^{\rm{H}}
\cdots{\bf{M}}_K^{\rm{H}}{\bf{Q}}_K^{\rm{H}}{\bf{Q}}_K{\bf{M}}_K
\cdots{\bf{Q}}_1{\bf{M}}_1\nonumber \\
& \ \ \ \ \ \ \ \ \ \ \ \ {\rm{Tr}}({\bf{F}}_{k}{\bf{F}}_k^{\rm{H}})\le P_k,  \ \ {\bf{Q}}_k^{\rm{H}}{\bf{Q}}_k={\bf{I}}_{N_{R,k}}.
\end{align}

Notice that the largest singular value of ${\bf{M}}_k$ is smaller than one. Therefore, the largest eigenvalue of ${\boldsymbol \Theta}$ is smaller than one (see Appendix~\ref{Appedix:5}) and then ${\bf{I}}_N-{\bf{Q}}_0^{\rm{H}}{\boldsymbol \Theta}{\bf{Q}}_0$ is a positive definite matrix. With the Cholesky factorization
\begin{align}
\label{L}
&({\bf{I}}_N-{\bf{Q}}_0^{\rm{H}}
{\boldsymbol \Theta}{\bf{Q}}_0)\sigma_b^2={\bf{L}}{\bf{L}}^{\rm{H}}
\end{align}where ${\bf{L}}$ is a lower triangular matrix, and the definition of ${\rm{MSE}}_n$ in the second line of (\ref{OPT_F_C}), we have
\begin{align}
\label{msn_n}
{\rm{MSE}}_n&=\sigma_b^2[{\bf{C}}({\bf{I}}_N-{\bf{Q}}_0^{\rm{H}}
{\boldsymbol \Theta}{\bf{Q}}_0){\bf{C}}^{\rm{H}}]_{n,n} \nonumber \\
&=([{\bf{C}}^{\rm{H}}]_{:,n})^{\rm{H}}{\bf{L}}{\bf{L}}^{\rm{H}}
[{\bf{C}}^{\rm{H}}]_{:,n} \nonumber \\
&=\sum_{i=1}^{n-1}[{\bf{L}}]_{i,i}^2|
[({\bf{C}}{\bf{L}}{\bf{D}}^{-1})^{\rm{H}}]_{i,n}|^2+[{\bf{L}}]_{n,n}^2\nonumber \\
& \ge [{\bf{L}}]_{n,n}^2,
\end{align}where ${\bf{D}}$ is a diagonal matrix defined as
\begin{align}
{\bf{D}}={\rm{diag}}\{[{\bf{L}}_{1,1}, \cdots, {\bf{L}}_{N,N}]^{\rm{T}}\}.
\end{align}In order to make the equality in the final line of (\ref{msn_n}) to hold, we need $
\sum_{i=1}^{n-1}[{\bf{L}}]_{i,i}^2|[({\bf{C}}{\bf{L}}{\bf{D}}^{-1})^{\rm{H}}]_{i,n}|^2$
$=0$, whose solution is
\begin{align}
\label{C_opt}
{\bf{C}}_{\rm{opt}}={\bf{D}}{\bf{L}}^{-1}.
\end{align}As a result $
{\rm{MSE}}_n=[{\bf{L}}]^2_{n,n}$, and the optimization problem for robust transceiver design is formulated as
\begin{align}
\label{OPT_F_0}
& \min_{{\bf{F}}_k,{\bf{Q}}_k} \ \ \ {{f}}({\left[[{\bf{L}}]^2_{1,1},\cdots,
[{\bf{L}}]^2_{N,N}\right]^{\rm{T}}}) \nonumber \\
& \ \  {\rm{s.t.}} \ \ \ \ \sigma_b^2({\bf{I}}_N-{\bf{Q}}_0^{\rm{H}}
{\boldsymbol \Theta}{\bf{Q}}_0)={\bf{L}}{\bf{L}}^{\rm{H}} \nonumber \\
& \ \ \ \ \ \ \ \ \ \  {\boldsymbol \Theta}={\bf{M}}_1^{\rm{H}}{\bf{Q}}_1^{\rm{H}}
\cdots{\bf{M}}_K^{\rm{H}}{\bf{Q}}_K^{\rm{H}}{\bf{Q}}_K{\bf{M}}_K
\cdots{\bf{Q}}_1{\bf{M}}_1\nonumber \\
& \ \ \ \ \ \ \ \ \ \
{\rm{Tr}}({\bf{F}}_{k}{\bf{F}}_k^{\rm{H}})\le P_k, \ \ {\bf{Q}}_k^{\rm{H}}{\bf{Q}}_k={\bf{I}}_{N_{R,k}}.
\end{align}


\section{Optimization Problem Reformulation for ${\bf{F}}_k$}

\subsection{Optimal Solution of ${\bf{Q}}_0$}
Because the objective function of the optimization problem (\ref{OPT_F_0}) is M-Schur-convex or M-Schur-concave. In the following, we will discuss the two cases separately.

\noindent {\textbf{\underline{M-Schur-convex:}}}

Taking the determinant on both sides of (\ref{L}), we have
\begin{align}
\label{determinant}
|\sigma_b^2({\bf{I}}_N-{\bf{Q}}_0^{\rm{H}}
{\boldsymbol \Theta}{\bf{Q}}_0)|=\prod_{n=1}^N[{\bf{L}}]_{n,n}^2=\sigma_b^{2N}\prod_{n=1}^N(1- \lambda_n({\boldsymbol \Theta}))
\end{align}where ${\lambda}_n({\boldsymbol
\Theta})$ is the $n^{\rm{th}}$ largest eigenvalue of ${\boldsymbol
\Theta}$. Based on (\ref{determinant}), the following multiplicative majorization relationship can be established \cite{Marshall79}
\begin{align}
\label{schur-convex-b}
\sigma_b^2\left[\prod_{n=1}^N(1- \lambda_n({\boldsymbol \Theta})) \right]^{\frac{1}{N}}\otimes{\bf{1}}_N\prec_{\times}{\left[[{\bf{L}}]^2_{1,1},\cdots,
[{\bf{L}}]^2_{N,N}\right]^{\rm{T}}},
\end{align}where the symbol $\otimes$ denotes
the Kronecker product and ${\bf{1}}_N$ is a $N\times 1$ all-one vector. With \textbf{Definition 2} and $f(\bullet)$ being a M-Schur-convex function, (\ref{schur-convex-b}) leads to
\begin{align}
\label{inequ_1} f({\left[[{\bf{L}}]^2_{1,1},\cdots, [{\bf{L}}]^2_{N,N}\right]^{\rm{T}}}) \!\ge \!
\underbrace{f \!\left(\!\sigma_b^2 \!\left[\prod_{n=1}^N(1 \!-\! \lambda_n({\boldsymbol \Theta}))\!
\right]^{\frac{1}{N}}\!\otimes\!{\bf{1}}_N \!\!\right)}_{\triangleq g[{\boldsymbol
\lambda}({\boldsymbol \Theta})]},
\end{align}where ${\boldsymbol \lambda}({\boldsymbol
\Theta})=[{\lambda}_1({\boldsymbol
\Theta}),\cdots,{\lambda}_N({\boldsymbol
\Theta})]^{\rm{T}}$.
The equality in (\ref{inequ_1}) holds when $\prec_{\times}$ in (\ref{schur-convex-b}) is replaced by equality, which means that $[{\bf{L}}]_{n,n}^2$ are identical for all $n$.  Notice that from (\ref{L}), we can write ${\bf{L}}{\bf{L}}^{\rm{H}}=\sigma_b^2{\bf{Q}}_0^{\rm{H}}({\bf{I}}-{\boldsymbol
\Theta}){\bf{Q}}_0$. Since ${\bf{I}}-{\boldsymbol
\Theta}$ is positive definite, there always exists an unitary matrix ${\bf{Q}}_{0}$ which makes the Cholesky factorization matrix of ${\bf{Q}}_0^{\rm{H}}({\bf{I}}-{\boldsymbol
\Theta}){\bf{Q}}_0$ have identical diagonal elements \cite{Amico2008}.
An explicit algorithm for constructing such ${\bf{Q}}_0$ is given in Appendix~\ref{Appedix:4}.

\noindent {\textbf{\underline{M-Schur-concave:}}}

From definition of ${\bf{L}}$ in (\ref{L}) and based Weyl' theorem \cite{Weyl1949}, we have
\begin{align}
\label{schur-a}
{\left[[{\bf{L}}]^2_{1,1},\cdots,
[{\bf{L}}]^2_{N,N}\right]^{\rm{T}}}
\prec_{\times}\sigma_b^2[{\bf{1}}_N-{\boldsymbol \lambda}({\boldsymbol
\Theta})].
\end{align}Applying $f(\bullet)$ on both sides of (\ref{schur-a}) and with \textbf{Definition 2}, we have
\begin{align}
\label{schur-convex-a}
f({\left[[{\bf{L}}]^2_{1,1},\cdots,
[{\bf{L}}]^2_{N,N}\right]^{\rm{T}}})\ge \underbrace{f(\sigma_b^2[{\bf{1}}_N-{\boldsymbol \lambda}({\boldsymbol
\Theta})])}_{\triangleq g[{\boldsymbol \lambda}({\boldsymbol
\Theta})]}.
\end{align}The equality in (\ref{schur-convex-a}) holds when
$\prec_{\times}$ in (\ref{schur-a}) is replaced by equality, which means that $[{\bf{L}}]_{n,n}^2$ equals to $\sigma_b^2[1-\lambda_n({\boldsymbol
\Theta})]$. On the other hand, taking eigenvalues on both sides of (\ref{L}), we can obtain $\sigma_b^2 [{\bf{1}}_N-{\boldsymbol \lambda}({\boldsymbol \Theta})]= [{\lambda}_N({\bf{L}}{\bf{L}}^{\rm{H}}),\cdots,{\lambda}_1({\bf{L}}{\bf{L}}^{\rm{H}})]^{\rm{T}}$. Therefore, $[[{\bf{L}}]_{1,1}^2,\cdots,[{\bf{L}}]_{N,N}^2]^{\rm{T}}=[{\lambda}_N({\bf{L}}{\bf{L}}^{\rm{H}}),\cdots,{\lambda}_1({\bf{L}}{\bf{L}}^{\rm{H}})]^{\rm{T}}$, which implies L is a diagonal matrix.  With ${\bf{L}}$ being a diagonal matrix,  ${\bf{Q}}_0^{\rm{H}}{\boldsymbol
\Theta}{\bf{Q}}_0$ is also a diagonal matrix. This can be satisfied if we take ${\bf{Q}}_0={\bf{U}}_{\boldsymbol \Theta}$, where the unitary matrix ${\bf{U}}_{\boldsymbol \Theta}$ is defined based on the eigendecomposition ${\boldsymbol \Theta}={\bf{U}}_{\boldsymbol \Theta}{\boldsymbol \Lambda}_{\boldsymbol \Theta}{\bf{U}}_{\boldsymbol \Theta}^{\rm{H}}$ with the elements of ${\boldsymbol \Lambda}_{\boldsymbol \Theta}$ arranged in decreasing order.


Notice that since ${\bf{L}}$ is a diagonal matrix, ${\bf{C}}_{\rm{opt}}$ in (\ref{C_opt}) is also a diagonal matrix.
Based on the definition of ${\bf{C}}$ in (\ref{C}) and with the fact that ${\bf{C}}$ is a lower triangular matrix with unit diagonal elements, it can be seen that the feedback matrix ${\bf{B}}$ must be an all-zero matrix. Therefore, when the objective function is M-Schur-concave, THP becomes linear precoding. The optimality of linear transceiver for M-Schur-concave objective function has also been obtained in point-to-point MIMO systems with perfect CSI \cite{Shenouda2008,Amico2008,Palomar2006}.

\noindent  {\textsl{\textbf{Remark 3:}}} The equal bit rate assumption at the beginning of Section~\ref{Section:signal_model} is for the operation of the nonlinear precoder only (this assumption also appears in \cite{Shenouda2008,Amico2008,Fischer2002}).  Notice that we have not used the equal bit rate assumption in the derivation of the optimal solution.  If the objective function is chosen such that a linear transceiver is obtained, this equal bit rate assumption will not appear in the solution.  On the other hand, if the objective function is chosen such that a nonlinear transceiver is obtained, the nature of the optimal transceiver is of equal bit rate (see the discussion below (\ref{inequ_1})).  Therefore, the equal bit rate assumption is not a restriction.


\noindent {\textbf{\underline{Summary:}}}

 Summarizing the previous results, when the objective function is M-Schur-convex or M-Schur-concave, the optimization problem (\ref{OPT_F_0}) is equivalent to
\begin{align}
\label{OPT_F_0_1}
& \min_{{\bf{F}}_k,{\bf{Q}}_k} \ \ \ {{g}}[{\boldsymbol \lambda}({\boldsymbol \Theta})] \nonumber \\
& \ \ {\rm{s.t.}} \ \ \ \ {\boldsymbol \Theta}={\bf{M}}_1^{\rm{H}}{\bf{Q}}_1^{\rm{H}}
\cdots{\bf{M}}_K^{\rm{H}}{\bf{Q}}_K^{\rm{H}}{\bf{Q}}_K{\bf{M}}_K
\cdots{\bf{Q}}_1{\bf{M}}_1\nonumber \\
&\ \ \ \ \ \ \ \ \ \ {\rm{Tr}}({\bf{F}}_{k}{\bf{F}}_k^{\rm{H}})\le P_k, \ \ {\bf{Q}}_k^{\rm{H}}{\bf{Q}}_k={\bf{I}}_{N_{R,k}}.
\end{align} where ${{g}}[{\boldsymbol \lambda}({\boldsymbol \Theta})]$ equals to
\begin{align}
\label{object_g} &{{g}}[{\boldsymbol \lambda}({\boldsymbol \Theta})]=
\begin{cases}
f(\sigma_b^2[\prod_{n=1}^N(1- \lambda_n({\boldsymbol \Theta}))]^{\frac{1}{N}}\otimes{\bf{1}}_N) \\
 \quad \quad\quad\quad\quad\quad\quad\quad\text{if $f(\bullet)$ is M-Schur-convex,}\\
f(\sigma_b^2[{\bf{1}}_N-{\boldsymbol \lambda}({\boldsymbol \Theta})])     \\
\quad\quad\quad\quad\quad\quad\quad\quad\text{if $f(\bullet)$ is M-Schur-concave.}\\
\end{cases}.
\end{align}

It is difficult to directly solve the optimization problem (\ref{OPT_F_0_1}), because ${\boldsymbol \Theta}$ is a product consists of matrices ${\bf{M}}_k$'s which in turn are
 complicated functions of the variables ${\bf{F}}_k$'s. In order to simplify the optimization problem (\ref{OPT_F_0_1}), we exploit the multiplicative majorization theory and transforms the objective function of (\ref{OPT_F_0_1}) to be a direct function of ${\bf{F}}_k$. To this end, we first provide useful results which form the theoretical basis of the following derivation.


\subsection{Prerequisites of Multiplicative Majorization Theory}



\noindent \textbf{Definition 3:}  For two vectors ${\bf{v}},{\bf{u}}\in \mathcal{D}$ with $\mathcal{D}= \{{\bf{z}}:z_1\ge \cdots \ge z_N \ge 0\}$, ${\bf{v}}\prec_{\times,w} {\bf{u}}$ is defined as
\begin{align}
\label{definition3}
& \prod_{i=1}^k {{v}}_{[i]}\le \prod_{i=1}^k {{u}}_{[i]}, \ \ k=1,\cdots,N.
\end{align}

Notice that there is a subtle difference between \textbf{Definition 2} in (\ref{def_2}) and \textbf{Definition 3}. In \textbf{Definition 3}, when $k=N$, $\prod_{i=1}^N {{v}}_{[i]}\le \prod_{i=1}^N {{u}}_{[i]}$ rather than $\prod_{i=1}^N {{v}}_{[i]}=\prod_{i=1}^N {{u}}_{[i]}$ in \textbf{Definition 2}.

\noindent \textbf{Lemma 2:} Let $\phi(\bullet)$ be a real-valued function on $\mathcal{D}$. Then $ \phi(\bullet)$ is decreasing and multiplicatively Schur-concave on $\mathcal{D}$ if and only if
 \begin{align}
 \label{Lemma_2_1}
 {\bf{v}}\prec_{\times,w} {\bf{u}} \Rightarrow {\phi}({\bf{v}})\ge { \phi}({\bf{u}}).
 \end{align} 

\noindent \textsl{\textbf{Proof:}} See Appendix~\ref{Appedix:2}. $\blacksquare$

\noindent \textbf{Lemma 3:} When $ \phi(\bullet)$ is increasing and multiplicatively Schur-concave, for ${\bf{v}},{\bf{u}} \in {\mathcal{C}}=\{{\bf{z}}:1> z_1\ge \cdots \ge z_N \ge 0\}$
\begin{align}
 {\bf{v}}\prec_{\times,w} {\bf{u}} \Rightarrow {\phi}({\bf{1}}_N-{\bf{v}})\ge { \phi}({\bf{1}}_N-{\bf{u}}).
 \end{align}

\noindent \textsl{\textbf{Proof:}} See Appendix~\ref{Appedix:3}. $\blacksquare$

%

\subsection{Problem Reformulation}

Based on the given results of multiplicative majorization theory, the optimization problem (\ref{OPT_F_0_1}) can be transformed into a much simpler one. Before presenting the result, two useful properties of the objective function $g(\bullet)$ are first derived based on the multiplicative majorization theory.

\noindent \textbf{Property 1:} The vector ${\boldsymbol \lambda}({\boldsymbol \Theta})$ has the following relationship
\begin{align}
&{\boldsymbol \lambda}({\boldsymbol \Theta}) \!\! \prec_{\times,w}\! \underbrace{[
\gamma_1(\!\{{\bf{F}}_k\}_{k=1}^K\!),\gamma_2(\!\{{\bf{F}}_k\}_{k=1}^K\!)
,\cdots,\gamma_N(\!\{{\bf{F}}_k\}_{k=1}^K\!) ]^{\rm{T}}}_{\triangleq {\boldsymbol
\gamma}(\{{\bf{F}}_k\}_{k=1}^K)}\nonumber \\ &\text{with}  \
{\gamma}_n(\{{\bf{F}}_k\}_{k=1}^K)=\prod_{k=1}^K{\frac{{ \lambda}_n({\bf{F}}_k^{\rm{H}}{\bf{\bar
H}}_k^{\rm{H}}{\bf{K}}_{{\bf{F}}_k}^{-1}{\bf{\bar H}}_k{\bf{F}}_k)}{1+{
\lambda}_n({\bf{F}}_k^{\rm{H}}{\bf{\bar H}}_k^{\rm{H}}{\bf{K}}_{{\bf{F}}_k}^{-1}{\bf{\bar
H}}_k{\bf{F}}_k)}},
\end{align}where the equality holds when
\begin{align}
\label{Q}
{\bf{Q}}_k={\bf{V}}_{{\bf{M}}_{k+1}}{\bf{U}}_{{\bf{M}}_{ k}}^{\rm{H}}, \ \ k=1,\cdots,K-1
\end{align}where ${\bf{U}}_{{\bf{M}}_k}$ and ${\bf{V}}_{{\bf{M}}_k}$ are defined based on the singular value decomposition ${\bf{M}}_k={\bf{U}}_{{\bf{M}}_k}{\boldsymbol \Lambda}_{{\bf{M}}_k}{\bf{V}}_{{\bf{M}}_k}^{\rm{H}}$ with the diagonal elements of ${\boldsymbol \Lambda}_{{\bf{M}}_k}$ arranged in decreasing order. Notice that (\ref{Q}) does not cover the design of ${\bf{Q}}_K$, but it can be any unitary matrix because it always appears in the form ${\bf{Q}}_K^{\rm{H}}{\bf{Q}}_K$ and equals to an identity matrix in the objective function.

\noindent \textsl{\textbf{Proof:}} See Appendix~\ref{Appedix:5}. $\blacksquare$

\noindent \textbf{Property 2:} The objective function $g[{\boldsymbol \lambda}({\boldsymbol
\Theta})]$ in (\ref{OPT_F_0_1}) is a decreasing M-Schur-concave function with respective to ${\boldsymbol \lambda}({\boldsymbol \Theta})$.

\noindent \textsl{\textbf{Proof:}} Based on \textbf{Lemma 2}, it is obvious that   $g[{\boldsymbol
\lambda}({\boldsymbol \Theta})]$ is a decreasing M-Schur-concave function if and only if ${\boldsymbol
\lambda}({\boldsymbol \Theta})\prec_{\times,w}  {{{\boldsymbol{\lambda}}({\boldsymbol{\tilde
\Theta}})}} \Rightarrow{g}[{\boldsymbol \lambda}({\boldsymbol \Theta})]\ge {g}[{{\boldsymbol{
\lambda}}({\boldsymbol{\tilde \Theta}})}]$. In the following, we will prove the latter.

When $f(\bullet)$ is M-Schur-convex, $g[{\boldsymbol \lambda}({\boldsymbol
\Theta})]=f(\sigma_b^2[\prod_{n=1}^N(1- \lambda_n({\boldsymbol \Theta}))]^{\frac{1}{N}}\otimes{\bf{1}}_N)$. Using \textbf{Lemma 1},  $\prod_{n=1}^N(1- \lambda_n({\boldsymbol \Theta}))$ can be proved to be a M-Schur-concave function of ${\boldsymbol \lambda}({\boldsymbol
\Theta})$. Furthermore, it can be easily seen that $\prod_{n=1}^N(1- \lambda_n({\boldsymbol \Theta}))$ is a decreasing function. If ${\boldsymbol \lambda}({\boldsymbol
\Theta})\prec_{\times,w} {{{\boldsymbol{ \lambda}}({\boldsymbol{\tilde
\Theta}})}}$ is true, based on \textbf{Lemma 2}, we have
\begin{align}
\prod_{n=1}^N(1- {\lambda}_n({\boldsymbol
\Theta}))\ge\prod_{n=1}^N(1-{\lambda}_n({\boldsymbol {\tilde{
\Theta}}})).
\end{align}Together with the fact that $f(\bullet)$ is an increasing function, it is concluded that
\begin{align}
&\underbrace{f(\sigma_b^2[\prod_{n=1}^N(1- {\lambda}_n({\boldsymbol \Theta}))]^{1/N}\otimes
{\bf{1}}_N)}_{g[{\boldsymbol {\lambda}}({\boldsymbol \Theta})]} \nonumber \\ &\ \ \ \ \  \  \ \ \ \ \
\ \ \ \ \ \ \ge \underbrace{f(\sigma_b^2[\prod_{n=1}^N(1- { \lambda}_n({\boldsymbol
{\tilde{\Theta}}}))]^{1/N}\otimes {\bf{1}}_N)}_{g[{\boldsymbol { {\lambda}}}({\boldsymbol {\tilde
\Theta}})]}.
\end{align}

On the other hand, when $f(\bullet)$ is increasing and M-Schur-concave, $g[{\boldsymbol \lambda}({\boldsymbol
\Theta})]=f(\sigma_b^2[{\bf{1}}_N-{\boldsymbol \lambda}({\boldsymbol
\Theta})])$. Using \textbf{Lemma 3} we directly have ${\boldsymbol \lambda}({\boldsymbol
\Theta})\prec_{\times,w} {\boldsymbol{ \lambda}}({\boldsymbol
{\tilde \Theta}})$ implies
\begin{align}
\underbrace{f(\sigma_b^2[{\bf{1}}_N-{\boldsymbol \lambda}({\boldsymbol
\Theta})])}_{g[{\boldsymbol { \lambda}}({\boldsymbol
\Theta})]}\ge \underbrace{f(\sigma_b^2[{\bf{1}}_N-{\boldsymbol { \lambda}}({\boldsymbol{\tilde
\Theta}})])}_{g[{\boldsymbol { \lambda}}({\boldsymbol
{\tilde \Theta}})]}.
\end{align} $\blacksquare$

Based on \textbf{Properties 1} and \textbf{2}, the objective function of (\ref{OPT_F_0_1}) has an achievable lower bound $g[{\boldsymbol \lambda}({\boldsymbol \Theta})] \ge g[{\boldsymbol \gamma}(\{{\bf{F}}_k\}_{k=1}^K)]$ with equality achieved when (\ref{Q}) is satisfied. When the lower bound is achieved, we have the following three additional observations:

\noindent (a) The constraints ${\bf{Q}}_k^{\rm{H}}{\bf{Q}}_k={\bf{I}}_{N_{R,k}}$ are automatically satisfied.

\noindent (b) The objective function $g[{\boldsymbol \gamma}(\{{\bf{F}}_k\}_{k=1}^K)]$ is independent of ${\bf{Q}}_k$.

\noindent (c) When ${\bf{F}}_k$'s are known, ${\bf{Q}}_k$'s can be directly computed using (\ref{Q}).

Applying these three observations into  (\ref{OPT_F_0_1}), we have the reformulated optimization problem
\begin{align}
\label{OPT_F_1}
& \min_{{\bf{F}}_k} \ \ \ {{g}}[{\boldsymbol \gamma}(\{{\bf{F}}_k\}_{k=1}^K)] \nonumber \\
& \ \ {\rm{s.t.}} \ \ \ \ \gamma_n(\{{\bf{F}}_k\}_{k=1}^K)=\prod_{k=1}^K\frac{{ \lambda}_n({\bf{F}}_k^{\rm{H}}{\bf{\bar H}}_k^{\rm{H}}{\bf{K}}_{{\bf{F}}_k}^{-1}{\bf{\bar H}}_k{\bf{F}}_k)}{1+{ \lambda}_n({\bf{F}}_k^{\rm{H}}{\bf{\bar H}}_k^{\rm{H}}{\bf{K}}_{{\bf{F}}_k}^{-1}{\bf{\bar H}}_k{\bf{F}}_k)} \nonumber \\
&\ \ \ \ \ \ \ \ \ \ {\rm{Tr}}({\bf{F}}_{k}{\bf{F}}_k^{\rm{H}})\le P_k.
\end{align}

\section{Solution of ${\bf{F}}_k$}

 In the following, we first derive the optimal structure of ${\bf{F}}_k$ and then present an algorithm to solve for the remaining unknown variables.


\subsection{Optimal Structure of ${\bf{F}}_k$}

Notice that ${{g}}(\bullet)$ is a decreasing function, and $ {\gamma}_n(\{{\bf{F}}_k\}_{k=1}^K)$ is an increasing function of $\lambda_n({\bf{ F}}_k^{\rm{H}}{\bf{\bar{H}}}_k^{\rm{H}}{\bf{K}}_{{\bf{F}}_k}^{-1}{\bf{\bar{H}}}_k
{\bf{ F}}_k)$. Therefore, ${{g}}[{\boldsymbol \gamma}(\{{\bf{F}}_k\}_{k=1}^K)]$ is a decreasing matrix-monotone function of ${\bf{ F}}_k^{\rm{H}}{\bf{\bar{H}}}_k^{\rm{H}}{\bf{K}}_{{\bf{F}}_k}^{-1}{\bf{\bar{H}}}_k
{\bf{F}}_k$ \cite{XingICSPCC}. Following the derivation in \cite{XingGlobecom}, it can be proved that at the optimal solution, the power constraints hold at the equality, i.e., ${\rm{Tr}}({\bf{F}}_k{\bf{F}}_k^{\rm{H}})=P_k$, meaning that the relays transmit at the maximum power.

Defining a variable $\eta_{f_k}$ as
 \begin{align}
 \label{eta_f}
 \eta_{f_k}&=\alpha_k{\rm{Tr}}({\bf{F}}_k{\bf{F}}_k^{\rm{H}}{\boldsymbol \Psi}_k)+\sigma_{n_k}^2 \ \ \text{with} \ \ \alpha_k={\rm{Tr}}({\boldsymbol \Sigma}_k)/N_{R,k},
\end{align} ${\rm{Tr}}({\bf{F}}_k{\bf{F}}_k^{\rm{H}})=P_k$ is exactly equivalent to ${\rm{Tr}}[{\bf{F}}_k
{\bf{F}}_k^{\rm{H}}(\alpha_k P_k{\boldsymbol \Psi}_k+\sigma_{n_k}^2{\bf{I}}_{N_{T,k}})]/\eta_{f_k}=P_k$ as proved in \cite{XingICSPCC,XingGlobecom,XingWCSP}. Thus the robust transceiver design problem (\ref{OPT_F_1}) is equivalent to
\begin{align}
\label{OPT_F_Final}
& \min_{{\bf{F}}_k} \ \ \ {{g}}[{\boldsymbol \gamma}(\{{\bf{F}}_k\}_{k=1}^K)] \nonumber \\
& \  {\rm{s.t.}} \ \ \ \  \gamma_n(\{{\bf{F}}_k\}_{k=1}^K)=\prod_{k=1}^K\frac{{ \lambda}_n({\bf{F}}_k^{\rm{H}}{\bf{\bar H}}_k^{\rm{H}}{\bf{K}}_{{\bf{F}}_k}^{-1}{\bf{\bar H}}_k{\bf{F}}_k)}{1+{ \lambda}_n({\bf{F}}_k^{\rm{H}}{\bf{\bar H}}_k^{\rm{H}}{\bf{K}}_{{\bf{F}}_k}^{-1}{\bf{\bar H}}_k{\bf{F}}_k)} \nonumber \\
&\  \ \ \ \ \ \ \ \ {\rm{Tr}}[{\bf{F}}_k
{\bf{F}}_k^{\rm{H}}(\alpha_k P_k{\boldsymbol \Psi}_k+\sigma_{n_k}^2{\bf{I}}_{N_{T,k}})]/\eta_{f_k}=P_k.
\end{align}It is proved in Appendix~\ref{Appedix:6} that  when ${\boldsymbol{ \Psi}}_k\propto {\bf{I}}_{N_{T,k}}$ or ${\boldsymbol \Sigma}_k\propto {\bf{I}}_{N_{R,k}}$, the optimal solutions of the optimization problem (\ref{OPT_F_Final}) have the following structure
\begin{align}
\label{Structure_F} {\bf{F}}_{k,\rm{opt}}&=\sqrt{{\xi}_k({\boldsymbol {
\Lambda}}_{{\boldsymbol{\mathcal{F}}}_k})}(\alpha_k P_k{\boldsymbol
\Psi}_k+\sigma_{n_k}^2{\bf{I}}_{N_{T,k}})^{-1/2}\nonumber \\
& \ \ \ \ \ \ \times {\bf{V}}_{{\boldsymbol{\mathcal{H}}}_k,N}
{\boldsymbol { \Lambda}}_{{\boldsymbol{\mathcal{F}}}_k}{\bf{U}}_{{\rm{Arb}}_{k},N}^{\rm{H}} \nonumber \\
{\text{with}} \ \ {\xi}_k({\boldsymbol { \Lambda}}_{{\boldsymbol{\mathcal{F}}}_k})
&={\sigma_{n_k}^2}/\{1-\alpha_k {\rm{Tr}}[{\bf{V}}_{{\boldsymbol{\mathcal{H}}}_k,N}^{\rm{H}}(\alpha_k
P_k{\boldsymbol
\Psi}_k +\sigma_{n_k}^2\nonumber \\
& \times  {\bf{I}}_{N_{T,k}})^{-1/2} {\boldsymbol \Psi}_k(\alpha_k P_k{\boldsymbol
\Psi}_k \!+ \! \sigma_{n_k}^2{\bf{I}}_{N_{T,k}})^{-1/2} \nonumber \\
&\ \ \ \ \ \ \ \ \  \ \ \ \  \ \ \ \  \ \ \ \ \ \ \ \times
{\bf{V}}_{{\boldsymbol{\mathcal{H}}}_k,N}{\boldsymbol{ \Lambda}}_{{\boldsymbol{\mathcal{F}}}_k}^2]\},
\end{align}where ${\boldsymbol{ \Lambda}}_{{\boldsymbol{\mathcal{F}}}_k}$ is a $N \times N$ unknown diagonal matrix, and  ${\bf{V}}_{{\boldsymbol{\mathcal{H}}}_k,N}$ and ${\bf{U}}_{{\rm{Arb}}_k,N}$ are the matrices consisting of the first $N$ columns of  ${\bf{V}}_{{\boldsymbol{\mathcal{H}}}_k}$ and ${\bf{U}}_{{\rm{Arb}}_k}$, respectively. The unitary matrix ${\bf{U}}_{{\rm{Arb}}_k}$ is an arbitrary $N_{R,k-1}\times N_{R,k-1}$ unitary matrix, and the unitary matrix ${\bf{V}}_{{\boldsymbol{\mathcal{H}}}_k}$ is defined based on the following singular value decomposition
\begin{align}
\label{decomp} & ({\bf{K}}_{{\bf{F}}_k}/\eta_{f_k})^{-1/2}{\bf{\bar H}}_k(\alpha_k P_k{\boldsymbol
\Psi}_k+{\sigma}_{n_k}^2{\bf{I}}_{N_{T,k}})^{-1/2} \nonumber \\
& \ \ \ \ \  \ \ \ \  \ \ \ \ \ \ \ \ \ \  \ \ \  \ \ \ \ \ \ \ \ \ \ \  \ \  \ \ \ \
={\bf{U}}_{{\boldsymbol{\mathcal{H}}}_k}{\boldsymbol
\Lambda}_{{\boldsymbol{\mathcal{H}}}_k}{\bf{V}}_{{\boldsymbol{\mathcal{H}}}_k}^{\rm{H}}
\end{align}where the diagonal elements of ${\boldsymbol \Lambda}_{{\boldsymbol{\mathcal{H}}}_k}$ are arranged in decreasing order.

\noindent {\textbf{\textsl{Remark 4:}}} In general, the expressions of
${\boldsymbol \Psi}_{k}$ and ${\boldsymbol \Sigma}_{k}$ depend on
specific channel estimation algorithms. Denote the transmit
and receive antennas correlation matrices and the channel estimation error variance in the $k^{\rm{th}}$ hop as ${\bf{R}}_{T,k}$, ${\bf{R}}_{R,k}$ and $ \sigma_{e,k}^2$, respectively. When the
channels are estimated based on the algorithm proposed in
\cite{Ding09,Ding10}, it can be shown that  $ {\boldsymbol{\Psi}}_{k}={\bf{R}}_{T,k}$ and
$
{\boldsymbol{\Sigma}}_{k}=\sigma_{e,k}^2({\bf{I}}_{N_{R,k}}+\sigma_{e,k}^2{\bf{R}}_{R,k}^{-1})^{-1}$.
 If the transmit antennas or the receive antennas are spaced widely,
 we have ${\bf{R}}_{T,k}\propto {\bf{I}}_{N_{T,k}}$ or
 ${\bf{R}}_{R,k}\propto {\bf{I}}_{N_{R,k}}$. These imply ${\boldsymbol{\Psi}}_{k}\propto {\bf{I}}_{N_{T,k}}$ or ${\boldsymbol{\Sigma}}_{k}\propto {\bf{I}}_{N_{R,k}}$.  Moreover, if the length of training is large, the value of $\sigma_{e,k}^2$ will be small and ${\bf{I}}_{N_{R,k}}+\sigma_{e,k}^2{\bf{R}}_{R,k}^{-1}\approx {\bf{I}}_{N_{R,k}}$. As a result,
${\boldsymbol{\Sigma}}_{k}$ will also approximate an identity matrix even when ${\bf{R}}_{R,k}\not\propto {\bf{I}}_{N_{R,k}}$. On the other hand, if the channel statistics are unknown, and using least-square channel estimator, it can be derived that ${\boldsymbol{\Sigma}}_{k}\propto {\bf{I}}_{N_{R,k}}$ always holds regardless of the antenna correlation or training length \cite{Xing1012}.

\subsection{Computation of ${\boldsymbol \Lambda}_{{\boldsymbol{\mathcal{F}}}_k}$}
It is obvious that in (\ref{Structure_F}), the only unknown variable is ${\boldsymbol \Lambda}_{{\boldsymbol{\mathcal{F}}}_k}$. In the following, we will discuss how to solve ${\boldsymbol \Lambda}_{{\boldsymbol{\mathcal{F}}}_k}$ in more detail. Denoting the following diagonal elements as
\begin{align}
[{\boldsymbol {\Lambda}}_{{\boldsymbol {\mathcal{H}}}_k}]_{n,n}=h_{k,n}, \ \  \ \
[{\boldsymbol {\Lambda}}_{{\boldsymbol{\mathcal{F}}}_k}]_{n,n}=f_{k,n},
\end{align}
substituting (\ref{Structure_F}) into the optimization problem (\ref{OPT_F_Final}) and noticing that ${\xi}_k({\boldsymbol { \Lambda}}_{{\boldsymbol{\mathcal{F}}}_k})=\eta_{f_k}$ (shown by (\ref{final}) in Appendix~\ref{Appedix:6}), after a straightforward derivation, the optimization for robust transceiver design is simplified as
\begin{align}
\label{Opt_F}
& \min_{f_{k,n}} \ \ \ {{g}}[{\boldsymbol \gamma}(\{{\bf{F}}_k\}_{k=1}^{K})] \nonumber \\
& \  {\rm{s.t.}} \ \ \ \  \gamma_n(\{{\bf{F}}_k\}_{k=1}^K)=\prod_{k=1}^K\frac{f_{k,n}^2{h}_{k,n}^2}{f_{k,n}^2{h}_{k,n}^2+1}\nonumber \\
&\ \ \ \ \ \ \ \ \sum_{n=1}^N f_{k,n}^2=P_k.
\end{align}The solution of (\ref{Opt_F}) depends on whether $f(\bullet)$ is M-Schur-convex or M-Schur-concave.

\noindent \underline{\textbf{M-Schur-convex functions:}}

Notice that when ${{f}}([{\rm{MSE}}_1,\cdots,{\rm{MSE}}_N]^{\rm{T}})$ is an M-Schur-convex function, regardless of the specific expression of $f(\bullet)$, the optimization problem (\ref{Opt_F}) is equivalent to minimize $\prod_{n=1}^N(1- \gamma_{n}(\{{\bf{F}}_k\}_{k=1}^K))$ \cite{Amico2008}. Therefore, the transceiver design problem (\ref{Opt_F}) equals to
\begin{align}
\label{M_Schur_convex}
& \min_{f_{k,i}} \ \ \ \sum_{n=1}^{N}{\rm{log}} \left(1- \frac{\prod_{k=1}^Kf_{k,n}^2h_{k,n}^2}{\prod_{k=1}^K(f_{k,n}^2h_{k,n}^2+1)}\right) \nonumber \\
& \ {\rm{s.t.}} \ \ \ \ \sum_{n=1}^N f_{k,n}^2 =P_k.
\end{align}In order to solve the optimization problem (\ref{M_Schur_convex}), iterative water-filling can be used to solve for $f_{k,i}$ with convergence guaranteed. More specifically, when $f_{l,i}$'s are fixed with $l\not=k$, $f_{k,i}$ is computed as
\begin{align}
\label{solution_capacity} & f_{k,n}^2=\frac{1}{h_{k,n}^2}\Bigg( \frac{-a_{k,n}+
\sqrt{a_{k,n}^2+4(1-a_{k,n})a_{k,n}h_{k,n}^2/\mu_k}}{2(1-a_{k,n})} \nonumber\\
&  \ \ \ \ \ \ \ \ \  \ \ \ \   \ \  \ \ \ \ \ \   -1 \Bigg)^{+} \ \ \  \ \ \ \ \ \  \ \  \ \  \ \ \ \ \ \ \   \ n=1,\cdots,N  \nonumber\\
& \text{with} \ \ a_{k,n}=\prod_{l\not=k}f_{l,n}^2h_{l,n}^2/(f_{l,n}^2h_{l,n}^2+1)
\end{align}where $\mu_k$ is the Lagrange multiplier which makes $\sum_{n=1}^N f_{k,n}^2 =P_k$ hold \cite{Boyd04}. Notice that this iterative water-filling algorithm is guaranteed to converge, as discussed in \cite{Yu04}.

\noindent \underline{\textbf{M-Schur-concave functions:}}

When ${{f}}([{\rm{MSE}}_1,\cdots,{\rm{MSE}}_N]^{\rm{T}})$ is a M-Schur-concave functions, there is no
unified solution. In this case, ${\boldsymbol \Lambda}_{{\boldsymbol{\mathcal{F}}}_k}$ should be
solved case by case. In the following, we use the example $
{{f}}([{\rm{MSE}}_1,\cdots,{\rm{MSE}}_N]^{\rm{T}})=\prod_{n=1}^N{\rm{MSE}}_n^{w_n}$ for $w_1\ge w_2
\cdots \ge w_N \ge 0$ to illustrate how to compute ${\boldsymbol
\Lambda}_{{\boldsymbol{\mathcal{F}}}_k}$. For this objective function, using (\ref{object_g}) it
follows that ${g} [{\boldsymbol \gamma}(\{{\bf{F}}_k\}_{k=1}^K)] =\sigma_{b}^{2{\sum}_n
w_n}\prod_{n=1}^{N}\left(1- {\gamma}_{n}(\{{\bf{F}}_k\}_{k=1}^K)\right)^{w_n}$ and the optimization
(\ref{Opt_F}) is equivalent to
\begin{align}
\label{46}
& \min_{f_{k,i}} \ \ \ \sum_{n=1}^{N}w_n{\rm{log}} \left(1- \frac{\prod_{k=1}^Kf_{k,n}^2h_{k,n}^2}{\prod_{k=1}^K(f_{k,n}^2h_{k,n}^2+1)}\right) \nonumber  \\
& \ {\rm{s.t.}} \ \ \ \sum_{n=1}^N f_{k,n}^2 =P_k.
\end{align}Equation (\ref{46}) has the same form as (\ref{M_Schur_convex}). Therefore, the solution can also be obtained by iterative water-filling solution.
Notice that the design problem becomes linear transceiver design problem when $f(\bullet)$ is M-Schur-concave.


\subsection{Summary and Implementation Issues}
 The design idea and procedure of the proposed robust transceiver are summarized in Table~\ref{Tabel:1}. For the implementation of the proposed algorithm, the execution  order is in reverse, i.e., from Step 8 to Step 3. Notice that in Step 8, iterative water-filling is adopted to solve for ${\boldsymbol \Lambda}_{{\boldsymbol {\mathcal{ F}}}_k}$. In general, only local optimality of the solution can be guaranteed \cite{Zhang2011}, which is a common problem for AF MIMO relaying design \cite{XingWCSP,Rong09,Mo09}.

\begin{table}
 \caption{Summary of Robust Transceiver Design}
 \centering
\begin{tabular}[t]{|c|l|}\hline \hline
1. & {{Derive the data estimation MSE matrix (\ref{MSE_0}).  }}\\\hline 2. & {{Formulate the
optimization problem (\ref{Problem_0}) with ${\bf{G}}$, ${\bf{P}}_k$ and ${\bf{C}}$}}
\\
&{{ as variables.}}\\ \hline
3. & {{Derive the optimal equalizer ${\bf{G}}$ as a function of ${\bf{P}}_k$ and ${\bf{C}}$  }}\\
& {{given by (\ref{G_opt}). Substitute the optimal ${\bf{G}}$ into the optimization }} \\
& {{problem (\ref{Problem_0}) to reduce the number of variables and have }}\\
&{{a reformulated optimization problem (\ref{Problem_1}). }}\\ \hline
4. & {{Simplify the constraints of the optimization problem (\ref{Problem_1})}}\\
& {{ by replacing ${\bf{P}}_k$ with ${\bf{Q}}_k$ and ${\bf{F}}_k$, and obtain an equivalent }}
\\
&{{optimization problem (\ref{OPT_F_C}).}}\\ \hline
5. &{{Derive the optimal ${\bf{C}}$ as a function of ${\bf{Q}}_k$ and ${\bf{F}}_k$ given }} \\
 & {{by (\ref{C_opt}). Substitute the optimal ${\bf{C}}$ into the optimization problem }}\\
 & {{(\ref{OPT_F_C}) and reformulate the optimization
  problem as (\ref{OPT_F_0}).}}\\\hline
6. & {{Derive the optimal ${\bf{Q}}_k$ as a function of ${\bf{F}}_k$ based on }} \\
 & {{majorization theory and substitute the optimal ${\bf{Q}}_k$ in (\ref{Q}) }}\\
 & {{into the optimization problem (\ref{OPT_F_0}) to reduce the number of }}\\
 &{variables. The optimization problem is then simplified to be (\ref{OPT_F_1}).}\\
 \hline
7. & {{Derive the optimal structure of ${\bf{F}}_k$ given by (\ref{Structure_F}).}} \\\hline 8.
&{{Solve for the unknown diagonal matrices }} ${\boldsymbol{\Lambda}}_{{\boldsymbol {\mathcal{F}}}_k}$
{{in the optimal }}\\ & {{structure using (\ref{Opt_F}).}}
\\\hline \hline
 \end{tabular}
 \label{Tabel:1}
\end{table}

 For information sharing in the implementation of the proposed solution, we can consider two algorithms.

\noindent \underline{\textbf{Central Algorithm:}}

 In centralized implementation, a natural assumption is that there is a central node performing the transceiver designs. All other nodes send its own estimated CSI to the central node via control channels, and after completing the design the central node informs each node the corresponding transceiver matrix. Since the channel does not change (or change very slowly), estimated CSI transmitted on control channels can be considered error-free due to low data transmission rates and heavy channel coding.


\noindent \underline{\textbf{Distributed Algorithm:}}

Based on the derived optimal structure ${\bf{F}}_{k,{\rm{opt}}}$ in (\ref{Structure_F}) and (\ref{decomp}) and the optimal ${\bf{Q}}_k$ in (\ref{Q}), using the definition of ${\bf{F}}_k$ given by (\ref{F_k_1}) and (\ref{F_k_2}), we can derive that the optimal forwarding matrix at the $k^{\rm{th}}$ node has the following structure
\begin{align}
\label{P_k_opt_a} {\bf{P}}_k &=(\alpha_k P_k{\boldsymbol
\Psi}_k+\sigma_{n_k}^2{\bf{I}}_{N_{T,k}})^{-1/2}\nonumber \\& \ \ \ \ \ \  \ \ \ \ \ \ \ \ \ \ \ \
\times {\bf{V}}_{{\boldsymbol{\mathcal{H}}}_k,N} {\boldsymbol {
\Lambda}}_{{\bf{P}}_k}{\bf{U}}_{{\boldsymbol{\mathcal{H}}}_{k-1},N}
^{\rm{H}}{\bf{K}}^{-1/2}_{{\bf{F}}_{k-1}},
\end{align}where ${\boldsymbol { \Lambda}}_{{\bf{P}}_k}$ is a diagonal matrix whose elements are functions of the diagonal elements of ${\boldsymbol \Lambda}_{{\boldsymbol {\mathcal F}}_m}$ for all $m$. It can be seen that except ${\boldsymbol { \Lambda}}_{{\bf{P}}_k}$, all other matrices in (\ref{P_k_opt_a}) are only the functions of the channels immediately preceding and succeeding the $k^{\rm{th}}$ node. It is easy for each node to obtain such channel information. As a result, the only information shared among all the other nodes is the diagonal elements of matrix ${\boldsymbol \Lambda}_{{\boldsymbol {\mathcal F}}_m}$ denoted by $\{f_{m,n}\}_{n=1}^N$. Notice that $\{f_{m,n}\}_{n=1}^N$ is the solution of the optimization problem (51). Exploiting the linear network topology and the fact that in the first constraint of (\ref{Opt_F}) $\{f_{m,n}\}_{n=1}^N$ appears in the form of $\prod_{k=1}^K{f_{k,n}^2{h}_{k,n}^2}/
({f_{k,n}^2{h}_{k,n}^2+1})$, only local information needs to be shared between adjacent nodes.

\begin{figure}[!t]
\centering
\includegraphics[width=.44\textwidth]{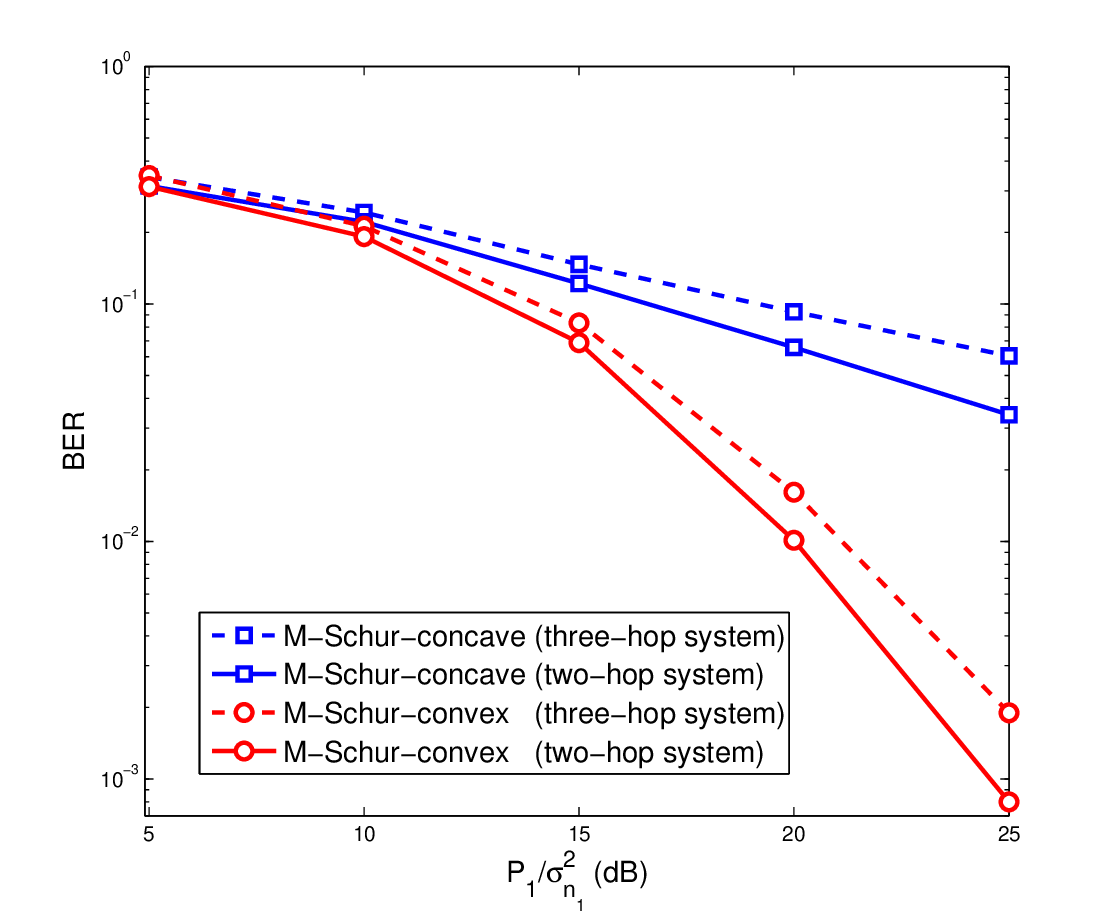}
\caption{BERs of the proposed transceivers with M-Schur-convex and M-Schur-concave objective functions
when $\rho_t=0$, $\rho_r=0.4$, $\sigma_e^2=0.001$.}\label{fig:2}
\end{figure}

\section{Simulation Results and Discussions}
In this section, the performance of the proposed algorithms is assessed by simulations. In the following, we consider an AF MIMO relaying
system where the source, relays and destination are all equipped with
four antennas, i.e., $N_{T,k}=N_{R,k}=4$. The estimation error
correlation matrices are chosen as the popular exponential model
$[{\boldsymbol \Psi}_k]=\sigma_e^2\rho_t^{|i-j|}$ and $[{\boldsymbol \Sigma}_k]=\rho_r^{|i-j|}$ \cite{Xing10}
where $\rho_t$ and $\rho_r$ are the correlation
coefficients, and $\sigma_{e}^2$ denotes the estimation error
variance. The estimated
channels ${\bf{\bar H}}_{k}$'s are randomly generated
 based on the following complex Gaussian distributions \cite{Xing10,Musavian07,Yoo2005}
\begin{align}
 &{\bf{\bar H}}_{k}\sim
\mathcal {C}\mathcal
{N}_{N_{R,k},N_{T,k}}({\bf{0}}_{N_{R,k},N_{T,k}},
\frac{(1-\sigma_{e}^2)}{\sigma_{e}^2}{\boldsymbol
\Sigma}_{k}
\otimes {\boldsymbol \Psi}_{k}^{\rm{T}}),
\end{align} such that channel realizations ${\bf{H}}_{k}={\bf{\bar
H}}_{k}+\Delta{\bf{H}}_{k}$ have unit variance. We define the signal-to-noise ratio (${\rm{SNR}}$) for the $k^{\rm{th}}$ link
 as $P_k/\sigma_{n_k}^2$. At the source node, four independent data streams are transmitted and in each data stream, ${N_{\rm{Data}}}=10000$ independent 16-QAM symbols are transmitted. Each point in the following figures is an average of 10000 trials and the bit error rates (BER) are computed \cite{Zhuhuiling2009,zhouyiqing2005,zhouyiqing2008}.

\begin{figure}[!t]
\centering
\includegraphics[width=.44\textwidth]{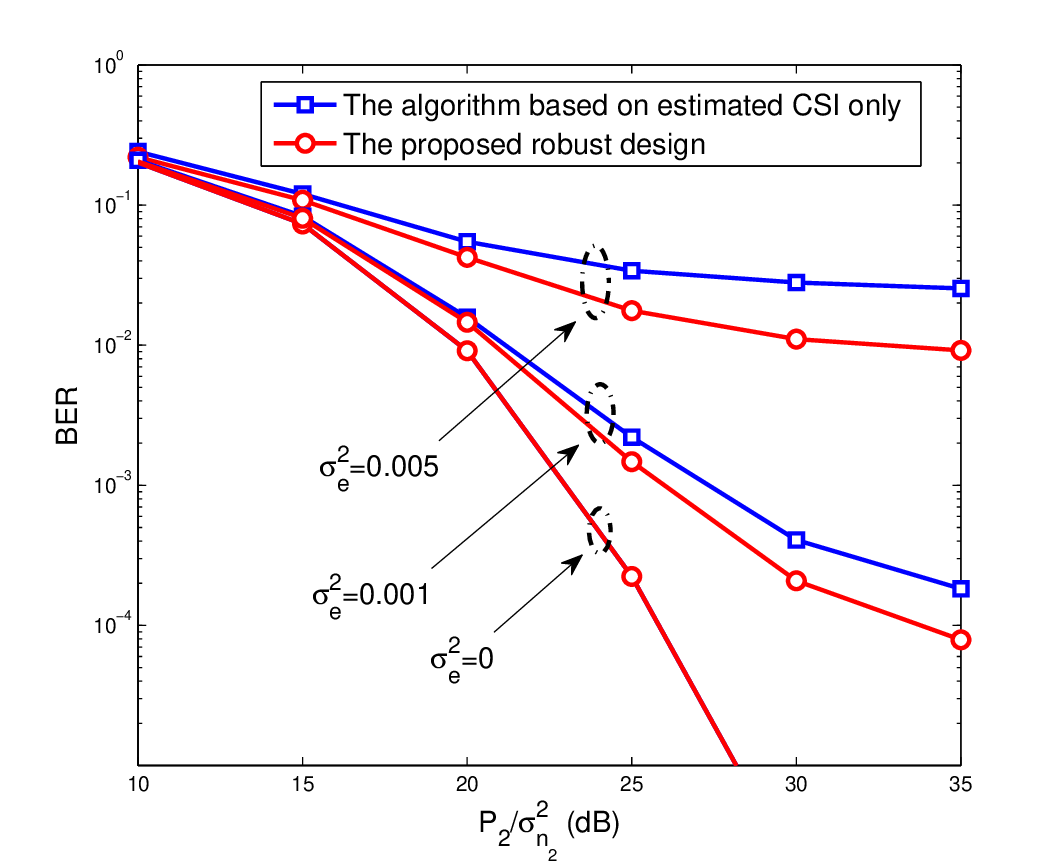}
\caption{BERs of proposed robust design with M-Schur-convex objective functions and the algorithm
based on estimated CSI only when $\rho_t=0.5$, $\rho_r=0$, and
$P_1/\sigma_{n_1}^2=30$dB.}\label{fig:3}
\end{figure}

First, we consider the objective function as weighted geometric mean MSE with equal weighting. In this case, the objective function is both M-Schur-convex and M-Schur-concave. There are two optimal solutions: one being linear transceiver and the other one being nonlinear transceiver. Fig.~\ref{fig:2} compares the BERs of these two solutions.
Both two-hop and three-hop systems are simulated with $\rho_t=0$, $\rho_r=0.4$, $\sigma_e^2=0.001$,  $P_2/\sigma_{n_2}^2=P_3/\sigma_{n_3}^2=30$dB and $P_1/\sigma_{n_1}^2$ being varied from 5 to 30dB. As expected, the nonlinear transceiver has a better performance than linear transceiver, but the performance improvement of nonlinear transceiver comes at the expense of higher complexity. Comparing to linear transceiver, the THP nonlinear transceiver has an additional $N\times N$ triangular matrix multiplication. Thus the additional complexity is $N(1+N)/2$ complex multiplications and $N(N-1)/2$ complex additions for each vector transmission. Furthermore, although the three-hop system performs not as good as the two-hop system, due to the extra hop of channel and noise amplification, the performance of the two-hop and three-hop systems shows the same trend. In the following, we focus on the M-Schur-convex objective function (i.e., nonlinear transceiver) for two-hop system only.


Next, we investigate the effect of the channel estimation error on the BER performance.
Fig.~\ref{fig:3} shows the BERs of the proposed robust nonlinear design and the corresponding
algorithm based on estimated CSI only (which takes the channel estimates as true channels) with
$\rho_t=0.5$, $\rho_t=0$, $P_1/\sigma_{n_1}^2=30$dB, and $P_2/\sigma_{n_2}^2$ being varied from 10 to
35dB. The algorithm based on estimated CSI only is obtained by simply setting
${\boldsymbol{\Psi}}_k={\bf{0}}$ in the proposed algorithm (similar approach has been used in
\cite{Zhang08} and \cite{Ding09}). From Fig.~\ref{fig:3}, it can be seen that smaller estimation
errors lead to better performance for both algorithms, but the performance of the proposed algorithm
is always better than that based on the estimated CSI only. Furthermore, the performance gap between
the proposed robust design and the algorithm based on estimated CSI becomes larger as the channel
estimation error increases. Of course, the performance of the two algorithms coincide when
$\sigma_e^2=0$.

\begin{figure}[!t]
\centering
\includegraphics[width=.44\textwidth]{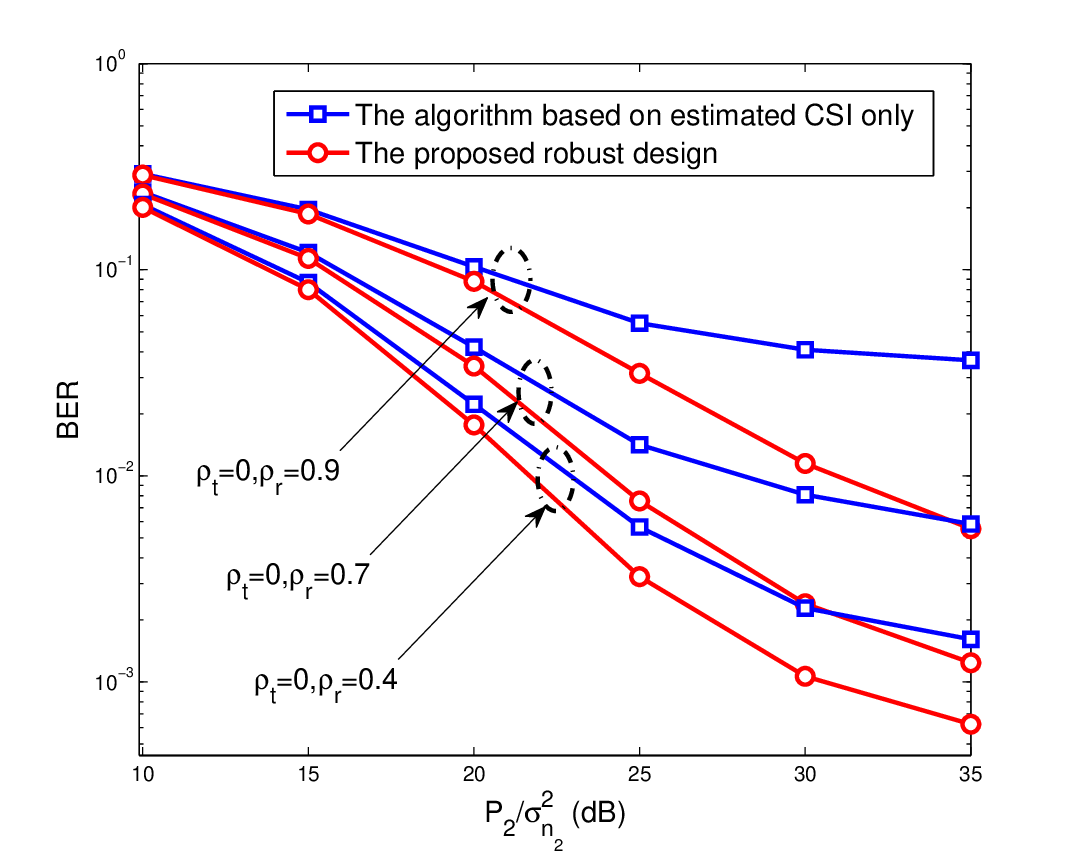}
\caption{BERs of proposed robust design with M-Schur-convex objective functions and the algorithm
based on estimated CSI only with different $\rho_r$, when $\rho_t=0$, $\sigma_e^2=0.002$ and
$P_1/\sigma_{n_1}^2=30$dB.}\label{fig:4}
\end{figure}

\begin{figure}[!t]
\centering
\includegraphics[width=.44\textwidth]{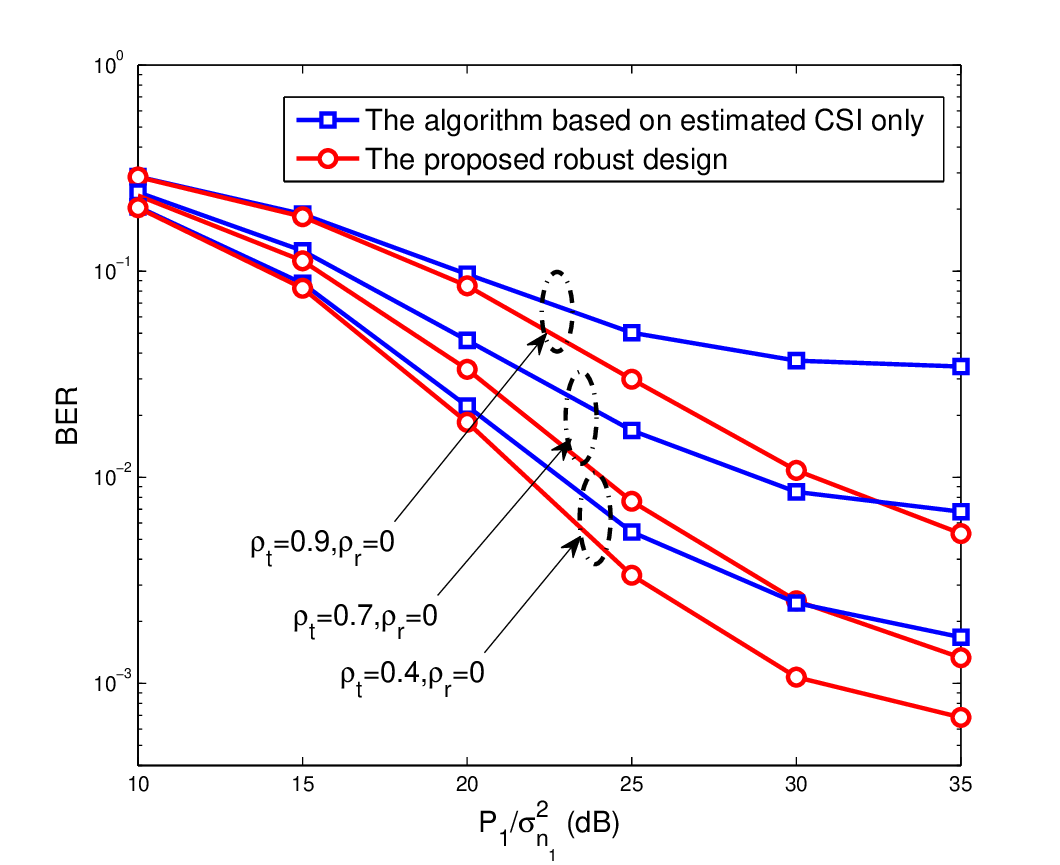}
\caption{BERs of proposed robust design with M-Schur-convex objective functions and the algorithm
based on estimated CSI only with different $\rho_t$, when $\rho_r=0$, $\sigma_e^2=0.002$ and
$P_2/\sigma_{n_2}^2=30$dB.}\label{fig:5}
\end{figure}

Finally, we illustrate the effects of correlation in the channel estimation errors. Fig.~\ref{fig:4} shows the BERs of the proposed robust design with M-Schur-convex objective functions and the corresponding algorithm based estimated CSI only for different $\rho_r$, when $\rho_t=0$, $\sigma_e^2=0.002$, $P_1/\sigma_{n_1}^2=30$dB, and $P_2/\sigma_{n_2}^2$ being varied from 10 to 35dB. It can be seen that in addition to the fact that the performance of the proposed robust design is always better than that based on the estimated CSI only, as $\rho_r$ increases, the performance gain of the proposed robust design with respect to that based on CSI only becomes larger.
It is most obvious when $\rho_r=0.9$ and at high SNR at the second hop. The performance gaps come from the fact that when correlation becomes stronger, ${\boldsymbol \Sigma}_k$ will be very different from identity matrix. Therefore from (\ref{Structure_F}) and ({\ref{decomp}}), the proposed optimal structure will be significantly different from that of the algorithm with estimated CSI only. As the designed precoding and forwarding matrices can be considered as the transmission directions, Figs.~\ref{fig:4} shows that correlation of channel estimation error would affect the direction of data transmission, and subsequently affect the final BER performance. Fig.~\ref{fig:5} shows the corresponding BERs for different $\rho_t$, with $\rho_r=0$, $\sigma_e^2=0.002$, $P_2/\sigma_{n_2}^2=30$dB, and $P_1/\sigma_{n_1}^2$ being varied from 10 to 35dB. It can be seen that a similar conclusion can be drawn.

\section{Conclusions}

Joint Bayesian robust transceiver design for multi-hop
 AF MIMO relaying systems was investigated. It was assumed that channel estimation errors exist in CSI in all hops. At the source node, a nonlinear Tomlinson-Harashima precoding was used, and was jointly optimized with linear forwarding matrices at all relays and linear equalizer at the destination. A general transceiver optimization problem was formulated with objective function being either M-Schur-convex or M-Schur-concave. Using elegant properties of multiplicative majorization theory and matrix-monotone functions, the optimal structure of the transceivers was first derived. Then, the original optimization problem was greatly simplified and an iterative water-filling solution was proposed to solve for the remaining unknown variables. Simulation results showed that the proposed robust design has much better performance than the non-robust design.

\appendices

\section{Proof of Lemma 1}
\label{Appedix:1}

Based on \textbf{Definition 2}, $\phi({\bf{z}})$ is M-Schur-convex over $\mathcal{D}=\{{\bf{z}}:z_1\ge \cdots \ge z_N \ge 0\}$ if and only if for ${\bf{v}},{\bf{u}}\in{\mathcal{D}}$, ${\bf{v}} \prec_{\times} {\bf{u}}$ implies $\phi({\bf{v}}) \le \phi({\bf{u}})$.

For a vector ${\bf{z}}\in \mathcal{D}$, define
\begin{align}
\label{tilde_z}
{\bf{\tilde z}}=[{{\tilde z}}_1,\cdots,{{\tilde z}}_N]^{\rm{T}} \ \ \text{and} \ \  {{\tilde z}}_k= \prod_{i=1}^k{z}_i.
\end{align} For ${\bf{v}},{\bf{u}}\in{\mathcal{D}}$, it is obvious that ${\bf{v}} \prec_{\times} {\bf{u}}$ is equivalent to
\begin{align}
\label{App_v}
& \{{{\tilde v}}_k\le {{\tilde u}}_k\}_{k=1}^{N-1}, \ \ \text{and} \ \ {{\tilde v}}_N= {{\tilde u}}_N.
\end{align}On the other hand, based on (\ref{tilde_z}), $z_k$ equals to
\begin{align}
\label{App_z}
z_k={\tilde z}_k/{\tilde z}_{k-1}, \ \ k\le L_z,
\end{align}where $L_z-1$ is the number of the nonzero elements of ${\bf{z}}$.
Therefore $\phi({\bf{v}})\le\phi({\bf{u}})$ can be written as
\begin{align}
\label{App_sz} &\underbrace{\phi({{\tilde v}}_1,{{\tilde v}}_2/{{\tilde v}}_1\cdots,{{\tilde
v}}_{L_v}/{{\tilde v}}_{L_v-1},0, \cdots,)}_{\triangleq \psi({\bf{\tilde v}})} \nonumber \\ & \ \ \ \
\ \ \ \ \ \le \underbrace{\phi({{\tilde u}}_1,{{\tilde u}}_2/{{\tilde u}}_1\cdots,{{\tilde
u}}_{L_u}/{{\tilde u}}_{L_u-1},0, \cdots,)}_{\triangleq \psi({\bf{\tilde u}})},
\end{align}

Based on (\ref{App_v}) and (\ref{App_sz}), proving $\phi({\bf{z}})$ is M-Schur-convex is equivalent to proving when $\{{{\tilde v}}_k\le {{\tilde u}}_k\}_{k=1}^{N-1}$ and ${{\tilde v}}_N= {{\tilde u}}_N$ hold, we have $\psi({\bf{\tilde v}}) \le \psi({\bf{\tilde u}})$. In other words, the proof becomes to prove $\psi(\bullet)$ is a vector-valued increasing function.

To prove $\psi(\bullet)$ is increasing, we only need to prove that when ${{\tilde v}}_k\le {{\tilde u}}_k$ and ${{\tilde v}}_l={{\tilde u}}_l$ for all $l{\not=}k$, we have $\psi( {\bf{\tilde v}}) \le \psi({\bf{\tilde u}})$ \cite{Marshall79}. As ${{\tilde v}}_k\ge0$ and ${{\tilde u}}_k\ge0$ , ${{\tilde v}}_k\le {{\tilde u}}_k$ is equivalent to ${{\tilde v}}_k={{\tilde u}}_k/e$ with $e\ge1$. Substituting ${{\tilde v}}_k={{\tilde u}}_k/e$ and ${{\tilde v}}_l={{\tilde u}}_l$ for all $l{\not=}k$ into (\ref{App_sz}) and replacing $u_k={\tilde u}_k/{\tilde u}_{k-1}$ for $k\le L_{u}-1$, proving $\psi( {\bf{\tilde v}}) \le \psi({\bf{\tilde u}})$ is equivalent to proving $\phi({{u}}_1,\cdots,{{u}}_k/e,{u}_{k+1}e, \cdots)$ is decreasing over $e\ge 1$ and ${{u}}_k/e\ge{u}_{k+1}e$.

\section{Algorithm for Computing ${\bf{Q}}_{0}$}
\label{Appedix:4}

 Following the sufficient conditions given in \cite{Zhang2005}, an explicit algorithm for constructing ${\bf{Q}}_0$ is given  as follows. Without loss of generality, in this Appendix, for both singular value decomposition (SVD) and eigendecomposition, the elements of the diagonal singular value or eigenvalue matrix are assumed to be in decreasing order.

\noindent \underline{\textbf{Step 1:}} Define ${\bf{A}}$ based on the following eigen-decomposition
\begin{align}
&({{\bf{I}}_N-
{\bf{M}}_1^{\rm{H}}{\bf{Q}}_1^{\rm{H}}\cdots{\bf{M}}_K^{\rm{H}}{\bf{Q}}_K^{\rm{H}}{\bf{Q}}_K{\bf{M}}_K\cdots{\bf{Q}}_1{\bf{M}}_1})^{1/2}\sigma_b
\nonumber \\ & \ \ \ \ \ \ \ \ \  \ \ \ \ \ \ \ \  \ \ \ \ \ \ \
 \ \ \ \  \ \ \ \ \ \ \ \ =\underbrace{{\bf{U}}_{{\bf{M}}}{\boldsymbol \Lambda}_{\bf{M}}}_{\triangleq
{\bf{A}}}{\bf{U}}_{{\bf{M}}}^{\rm{H}}.
\end{align}

\noindent \underline{\textbf{Step 2:}} Initialize ${\bf{S}}={\bf{0}}_{N\times N}$ and set
\begin{align}
[{\bf{S}}]_{1,1}=&\sqrt {\frac{|{\bf{A}}^{\rm{H}}{\bf{A}}|^{1/N}-[{\boldsymbol
\Lambda}_{\bf{M}}]_{N,N}}{[{\boldsymbol \Lambda}_{\bf{M}}]_{1,1}-[{\boldsymbol
\Lambda}_{\bf{M}}]_{N,N}}}, \nonumber \\ [{\bf{S}}]_{N,1}=& \sqrt{\frac{[{\boldsymbol
\Lambda}_{\bf{M}}]_{1,1}-|{\bf{A}}^{\rm{H}}{\bf{A}}|^{1/N}}{[{\boldsymbol
\Lambda}_{\bf{M}}]_{1,1}-[{\boldsymbol \Lambda}_{\bf{M}}]_{N,N}}}.
\end{align}Meanwhile, the orthogonal complement matrix of $[{\bf{S}}]_{:,1}$ is set to be
\begin{align}
[{\bf{S}}]_{:,1}^{\bot}=
\left[ {\begin{array}{*{20}c}
   {-[{\bf{S}}]_{N,1}} & {{\bf{0}}}  \\
   {{\bf{0}}} & {{\bf{I}}}  \\
   {[{\bf{S}}]_{1,1}} & {{\bf{0}}}  \\
\end{array}} \right].
\end{align}
\noindent \underline{\textbf{Step 3:}} Begin recursion for $k=1,\cdots,N-2$.
Compute a $(N-k)\times (N-k)$ unitary matrix ${\bf{V}}^{(k)}$ based on the following eigendecomposition
\begin{align}
&({\bf{A}}[{\bf{S}}]_{:,1:k}^{\bot})^{\rm{H}} [{\bf{I}}-{\bf{A}}[{\bf{S}}]_{:,1:k}
([{\bf{S}}]_{:,1:k}^{\rm{H}}{\bf{A}}^{\rm{H}}{\bf{A}} [{\bf{S}}]_{:,1:k})^{-1}\nonumber \\& \ \ \ \ \
\times [{\bf{S}}]_{:,1:k}^{\rm{H}}{\bf{A}}^{\rm{H}}] ({\bf{A}}[{\bf{S}}]_{:,1:k}^{\bot})
={\bf{V}}^{(k)}{\boldsymbol \Lambda}^{(k)}({\bf{V}}^{(k)})^{\rm{H}}.
\end{align}Then update the $(k+1)^{\rm{th}}$ column of ${\bf{S}}$ as
\begin{align}
[{\bf{S}}]_{:,k+1}=[{\bf{S}}]_{:,1:k}^{\bot}{\bf{V}}^{(k)}{\bf{y}}^{(k)}
\end{align} and
\begin{align}
{\bf{y}}^{(k)} &=\Bigg[\sqrt{\frac{|{\bf{A}}^{\rm{H}}{\bf{A}}|^{1/N}-[{\boldsymbol
\Lambda}^{(k)}]_{N-k,N-k}} {[{\boldsymbol \Lambda}^{(k)}]_{1,1}-[{\boldsymbol
\Lambda}^{(k)}]_{N-k,N-k}}}, {\bf{0}}_{1,N-k-1}, \nonumber \\ & \ \ \ \ \ \ \ \ \  \ \ \ \ \  \ \ \ \
\ \sqrt{\frac{[{\boldsymbol \Lambda}^{(k)}]_{1,1} -|{\bf{A}}^{\rm{H}}{\bf{A}}|^{1/N}}{[{\boldsymbol
\Lambda}^{(k)}]_{1,1}-[{\boldsymbol \Lambda}^{(k)}]_{N-k,N-k}}}\Bigg]^{\rm{T}}.
\end{align} Based on the SVD ${\bf{S}}={\bf{U}}_{\bf{S}}{\boldsymbol \Lambda}_{\bf{S}}{\bf{V}}_{\bf{S}}^{\rm{H}}$, the orthogonal complement matrix of $[{\bf{S}}]_{:,1:k+1}$ is computed as
\begin{align}
[{\bf{S}}]_{:,1:k+1}^{\bot}=[{\bf{U}}_{\bf{S}}]_{:,k+2:N}.
\end{align}

\noindent \underline{\textbf{Step 4:}}
When $k=N-1$, $
[{\bf{S}}]_{:,N}=[{\bf{S}}]_{:,1:N-2}^{\bot}{\bf{V}}^{(N-2)}{\bf{y}}^{(N-1)}$ and
\begin{align}
{\bf{y}}^{(N-1)}&=\Bigg[\sqrt{\frac{[{\boldsymbol \Lambda}^{(N-2)}]_{1,1}
-|{\bf{A}}^{\rm{H}}{\bf{A}}|^{1/N}}{[{\boldsymbol \Lambda}^{(N-2)}]_{1,1}-[{\boldsymbol
\Lambda}^{(N-2)}]_{2,2}}},\nonumber \\& \ \ \ \ \ \ \ \ \ \ \ \
-\sqrt{\frac{|{\bf{A}}^{\rm{H}}{\bf{A}}|^{1/N}-[{\boldsymbol \Lambda}^{(N-2)}]_{2,2}} {[{\boldsymbol
\Lambda}^{(N-2)}]_{1,1}-[{\boldsymbol \Lambda}^{(N-2)}]_{2,2}}}\Bigg]^{\rm{T}}.
\end{align}
\noindent \underline{\textbf{Step 5:}} Finally, ${\bf{Q}}_{0}$ equals to ${\bf{Q}}_{0}={\bf{U}}_{\bf{M}}{\bf{S}}$.

\section{Proof of Lemma 2}
\label{Appedix:2}

\noindent \underline{\textbf{Proof of ``if'' direction}}

First, we will prove that for any two vectors ${\bf{v}},{\bf{u}}\in{\mathcal{D}}$, ${\bf{v}} \prec_{\times,w} {\bf{u}}\Rightarrow \phi({\bf{v}}) \ge \phi({\bf{u}})$ implies $\phi(\bullet)$  is a decreasing M-Schur-concave function over ${\mathcal{D}}$.

When ${\bf{v}}\prec_{\times,w}{\bf{u}}\Rightarrow \phi({\bf{v}})\ge \phi({\bf{u}})$ holds, ${\bf{v}}\prec_{\times}{\bf{u}}\Rightarrow \phi({\bf{v}})\ge \phi({\bf{u}})$ must hold. Using \textbf{Lemma 1}, $\phi(\bullet)$ must be M-Schur-concave over ${\mathcal{D}}$.

Furthermore, for ${\bf{v}},{\bf{u}} \in \mathcal{D}$ with $v_k\le u_k$ and $v_i=u_i$ for all $i{\not=}k$, we have  ${\bf{v}} \prec_{\times,w} {\bf{u}}$.  Then ${\bf{v}} \prec_{\times,w} {\bf{u}}\Rightarrow \phi({\bf{v}}) \ge \phi({\bf{u}})$ implies $\phi(\bullet)$ is a decreasing function. Therefore, when ${\bf{v}} \prec_{\times,w} {\bf{u}}\Rightarrow \phi({\bf{v}}) \ge \phi({\bf{u}})$, then we have $\phi(\bullet)$ is a decreasing M-Schur-concave function.

\noindent \underline{\textbf{Proof of ``only if'' direction}}

On the other hand, when $\phi(\bullet)$ is a decreasing M-Schur-concave function, we need prove that ${\bf{v}} \prec_{\times,w} {\bf{u}}\Rightarrow \phi({\bf{v}}) \ge \phi({\bf{u}})$. For any two vectors ${\bf{v}},{\bf{u}} \in\mathcal{D}$ with ${\bf{v}} \prec_{\times,w} {\bf{u}}$  we can construct a vector ${\boldsymbol{\tau}} \in \mathcal{D}$ with $\tau_i=u_i$ for $i<N$ and $\tau_N$ is chosen to makes $\prod_{i=1}^N\tau_i=\prod_{i=1}^Nv_i$. It is obvious that $\tau_N\le u_N$. Then if ${\bf{v}}\prec_{\times,w}{\bf{u}}$, we have  ${\bf{v}}\prec_{\times}{\boldsymbol{\tau}}$ and ${\boldsymbol{\tau}}\prec_{\times,w}{\bf{u}}$.

As $\phi(\bullet)$ is M-Schur-concave, based on \textbf{Lemma 1} we directly have $\phi({\bf{v}})\ge\phi({\boldsymbol{\tau}})$. Furthermore, since the difference between ${\boldsymbol \tau}$ and ${\bf{u}}$ is only in the last element with $\tau_N\le u_N$, as $\phi(\bullet)$ is decreasing, we have $\phi({\boldsymbol{\tau}})\ge \phi({\bf{u}})$. Combining the two inequalities, we have $\phi({\bf{v}})\le\phi({\bf{u}})$.

\section{Proof of Lemma 3}
\label{Appedix:3}
Based on \textbf{Lemma 1}, it can be proved that $\prod_{i=1}^k(1-{z}_i)$ is an M-Schur-concave function. It is also obvious that $\prod_{i=1}^k(1-{z}_i)$ is a decreasing function for ${\bf{z}} \in {\mathcal{C}}=\{{\bf{z}}:1> z_1\ge \cdots \ge z_N \ge 0\}$. Using \textbf{Lemma 2}, for ${\bf{v}},{\bf{u}}\in {\mathcal{C}}$ with ${\bf{v}} \prec_{\Pi} {\bf{u}}$, we have
\begin{align}
\label{v_u}
\prod_{i=1}^k\underbrace{(1-{v}_i)}_{\triangleq {\hat v}_{(i)}}\ge \prod_{i=1}^k\underbrace{(1-{u}_i)}_{\triangleq {\hat u}_{(i)}}>0, \ \ k=1,\cdots,N.
\end{align}

We construct a vector ${\boldsymbol{\hat \tau}}=[{{\hat \tau}}_{(1)},\cdots,{{\hat \tau}}_{(N)}]^{\rm{T}}$ with ${{\hat \tau}}_{(i)}={{\hat u}}_{(i)}$
for $i<N$ and ${\hat \tau}_{(N)}$ is chosen to makes $\prod_{i=1}^N{\hat v}_{(i)}=\prod_{i=1}^N{\hat \tau}_{(i)}$ hold. It is obvious that ${\tau}_{(N)}\ge {u}_{(N)}$. As the only difference between ${\tau}_{(i)}$ and ${u}_{(i)}$ is at $i=N$, when $\phi(\bullet)$ is increasing, we have $ {\phi}({\boldsymbol{\hat \tau}})\ge {\phi}({\bf{\hat u}})$ where ${\bf{\hat u}}=[{{\hat u}}_{(1)},\cdots,{{\hat u}}_{(N)}]^{\rm{T}}$.

On the other hand, based on (\ref{v_u}) and the fact that ${{\hat \tau}}_{(i)}={{\hat u}}_{(i)}$
for $i<N$, it can be concluded that (a) $\prod_{i=1}^k{\hat v}_{(i)}\ge \prod_{i=1}^k{\hat \tau}_{(i)}$ for $1\le k<N$.  Based on the definition of ${\hat \tau}_N$, it can also be concluded that (b) $\prod_{i=1}^N{\hat v}_{(i)}= \prod_{i=1}^N{\hat \tau}_{(i)}>0$. Results (a) and (b) implies ${\bf{\hat v}} \prec_{\times} {\boldsymbol{\hat \tau}}$ where ${\bf{\hat v}}=[{{\hat v}}_{(1)},\cdots,{{\hat v}}_{(N)}]^{\rm{T}}$ \cite{Marshall79}. As $\phi(\bullet)$ M-Schur-concave, using \textbf{Lemma 1}, we have ${\phi}({\bf{\hat v}})\ge {\phi}({\boldsymbol{\hat \tau}})$.
Together with the conclusion in the last paragraph, we can obtain ${\phi}({\bf{\hat v}})\ge {\phi}({\bf{\hat u}})$. Finally, with ${\bf{\hat v}}={\bf{1}}_N-{\bf{v}}$ and ${\bf{\hat u}}={\bf{1}}_N-{\bf{u}}$, the proof is completed.

\section{Proof of Property 1}
\label{Appedix:5}
First notice two facts in matrix theory: (a) for two matrices ${\boldsymbol A}$ and ${\boldsymbol B}$ with compatible dimension
$\lambda_i({\boldsymbol A}{\boldsymbol B})=\lambda_i({\boldsymbol B}{\boldsymbol A})$ \cite[9.A.1.a]{Marshall79}; (b) for two positive semi-definite matrices ${\boldsymbol A}$ and ${\boldsymbol B}$, $\prod_{i=1}^{n}\lambda_{i}({\boldsymbol A}{\boldsymbol B})\le\prod_{i=1}^{n}{\lambda_{i}({\boldsymbol A})\lambda_{i}({\boldsymbol B})}$ \cite[9.H.1.a]{Marshall79}, where the equality holds when ${\boldsymbol A}$ and ${\boldsymbol B}$ has the same unitary matrix in eigendecomposition. With these two facts, we have
\begin{align}
& \prod_{i=1}^{n}\lambda_{i}({{\bf M}_1^{\rm{H}}{\bf{Q}}_{1}^{\rm{H}}}{\cdots {\bf M}_K^{\rm{H}}{\bf{Q}}_{K}^{\rm{H}}{\bf{Q}}_{K} {\bf M}_K\cdots {\bf{Q}}_{1}{\bf M}_{1}}) \nonumber \\
=& \prod_{i=1}^{n}\lambda_{i}({{\bf M}_{2}^{\rm{H}}{\bf{Q}}_{2}^{\rm{H}}\cdots {\bf M}_K^{\rm{H}}{\bf{Q}}_{K}^{\rm{H}}{\bf{Q}}_{K} {\bf M}_K \cdots} \nonumber \\
& \ \ \ \ \ \ \  \ \ \ \  \ \  \ \  \ \  \ \ \ \ \ \ \ \ \times {\bf{Q}}_{2}{\bf M}_{2} {{\bf{Q}}_{1}{\bf M}_{1}{\bf M}_1^{\rm{H}}{\bf{Q}}_{1}^{\rm{H}}}) \nonumber \\
\le & \prod_{i=1}^{n}[\lambda_i({\bf M}_2^{\rm{H}}{\bf{Q}}_{2}^{\rm{H}}\cdots {\bf
M}_K^{\rm{H}}{\bf{Q}}_{K}^{\rm{H}}{\bf{Q}}_{K} {\bf M}_K\cdots {\bf{Q}}_{2}{\bf M}_{2})\nonumber \\
&  \ \ \ \times \lambda_{i}({\bf M}_1{\bf M}_1^{\rm{H}})]  \ \ \ \ \ \ \ \ \ n=1,\cdots,N,
\end{align}where the first equality is due to fact (a) and the second inequality is based on fact (b). Repeating the above two processes and based on the fact that $\lambda_i({\bf M}_k{\bf M}_k^{\rm{H}})=\lambda_i({\bf M}_k^{\rm{H}}{\bf M}_k)$  we can obtain the following inequality
\begin{align}
\label{App_Inequality} & \ \ \ \ \prod_{i=1}^{n}\lambda_{i}({\boldsymbol \Theta}) \nonumber \\
&\le \prod_{i=1}^{n}[\underbrace{\lambda_{i}({\bf M}_K^{\rm{H}}{\bf M}_K)\lambda_i({\bf
M}_{K-1}^{\rm{H}}{\bf M}_{K-1})\cdots \lambda_i({\bf M}_1^{\rm{H}}{\bf M}_1)}_{\triangleq
{\gamma}_{i}(\{{\bf{F}}_k\}_{k=1}^K)}],
\end{align}where the equality holds when ${\bf{Q}}_k$'s satisfy
\begin{align}
\label{72} {\bf{Q}}_k={\bf{V}}_{{\bf{M}}_{k+1}}{\bf{U}}_{{\bf{M}}_{k}}^{\rm{H}}, \ \ k=1,\cdots,K-1,
\end{align}where ${\bf{U}}_{{\bf{M}}_k}$ and ${\bf{V}}_{{\bf{M}}_{k}}$ are defined based on the following singular value decomposition ${\bf{M}}_k={\bf{U}}_{{\bf{M}}_k}
{\boldsymbol{\Lambda}}_{{\bf{M}}_k}{\bf{V}}_{{\bf{M}}_k}^{\rm{H}}$ with the diagonal elements of  ${\boldsymbol{\Lambda}}_{{\bf{M}}_k}$ arranged in decreasing order. Furthermore, based on the definition of ${\bf M}_k$ in (\ref{object_MSE}), $\gamma_i(\{{\bf{F}}_k\}_{k=1}^K)$ in (\ref{App_Inequality}) equals to
\begin{align}
\gamma_i(\{{\bf{F}}_k\}_{k=1}^K)=\prod_{k=1}^K\frac{{ \lambda}_i({\bf{F}}_k^{\rm{H}}{\bf{\bar H}}_k^{\rm{H}}{\bf{K}}_{{\bf{F}}_k}^{-1}{\bf{\bar H}}_k{\bf{F}}_k)}{1+{ \lambda}_i({\bf{F}}_k^{\rm{H}}{\bf{\bar H}}_k^{\rm{H}}{\bf{K}}_{{\bf{F}}_k}^{-1}{\bf{\bar H}}_k{\bf{F}}_k)}.
\end{align}


\section{Optimal Structure of ${\bf{F}}_k$}
\label{Appedix:6}

Defining new variables
\begin{align}
{\bf{\tilde F}}_k&={1}/{\sqrt{\eta_{f_k}}}(\alpha_k P_k {\boldsymbol
\Psi}_k+\sigma_{n_k}^2{\bf{I}}_{N_{T,k}})^{1/2}{\bf{F}}_k, \ \ \text{and}  \nonumber \\
{\boldsymbol {\mathcal{H}}}_k&=({\bf{K}}_{{\bf{F}}_k}/\eta_{f_k})^{-1/2}{\bf{\bar H}}_k(\alpha_k P_k{\boldsymbol \Psi}_k+{\sigma}_{n_k}^2{\bf{I}}_{N_{T,k}})^{-1/2},
\end{align}
the optimization problem (\ref{OPT_F_Final}) is reformulated as
\begin{align}
\label{OPT_F_2}
& \min_{{\bf{\tilde F}}_k} \ \ \ {{g}}[{\boldsymbol \gamma}(\{{\bf{\tilde F}}_k\}_{k=1}^K)] \nonumber \\
& \ \ {\rm{s.t.}} \ \ \ \gamma_n(\{{\bf{\tilde F}}_k\}_{k=1}^K)=\prod_{k=1}^K\frac{{ \lambda}_n({\bf{\tilde F}}_k^{\rm{H}}{\boldsymbol{\mathcal {H}}}_k^{\rm{H}}{\boldsymbol{\mathcal { H}}}_k{\bf{\tilde F}}_k)}{1+{ \lambda}_n({\bf{\tilde F}}_k^{\rm{H}}{\boldsymbol{\mathcal {H}}}_k^{\rm{H}}{\boldsymbol{\mathcal { H}}}_k{\bf{\tilde F}}_k)} \nonumber \\
&\ \ \ \ \ \ \ \ \ {\rm{Tr}}({\bf{\tilde F}}_{k}{\bf{\tilde F}}_k^{\rm{H}})= P_k.
\end{align}
When ${\boldsymbol{ \Psi}}_k\propto {\bf{I}}_{N_{T,k}}$ or ${\boldsymbol \Sigma}_k\propto {\bf{I}}_{N_{R,k}}$, for the optimal solution ${\bf{K}}_{{\bf{F}}_k}/\eta_{f_k}$ is constant \cite{XingICSPCC,XingGlobecom,XingWCSP} and thus ${\boldsymbol {\mathcal{H}}}_k$
is constant. Let ${\bf{\tilde F}}_{k,{\rm{opt}}}$ be the optimal solution of (\ref{OPT_F_2}). With the following singular value decompositions,
\begin{align}
&{\boldsymbol {\mathcal{H}}}_k{\bf{\tilde F}}_{k,{\rm{opt}}}={\bf{U}}_{{\boldsymbol A}_{k}}
{\boldsymbol \Lambda}_{{\boldsymbol A}_{k}}{\bf{V}}_{{\boldsymbol A}_{k}}^{\rm{H}},  \ \ \
{\boldsymbol {\mathcal{H}}}_k={\bf{U}}_{{\boldsymbol {\mathcal{H}}}_k}{\boldsymbol
\Lambda}_{{\boldsymbol {\mathcal{H}}}_k}{\bf{V}}_{{\boldsymbol {\mathcal{H}}}_k}^{\rm{H}},
\end{align}where the diagonal elements of ${\boldsymbol \Lambda}_{{\boldsymbol A}_{k}}$ and ${\boldsymbol \Lambda}_{{\boldsymbol {\mathcal{H}}}_k}$ are arranged  in decreasing order, we can construct a matrix ${\bf{\hat F}}_k$ equals to
\begin{align}
\label{Apped_F} {\bf{\hat { F}}}_k={\bf{V}}_{{\boldsymbol {\mathcal{H}}}_k}{\boldsymbol
\Lambda}_{{\bf{X}}_k}{\bf{V}}_{{\boldsymbol A}_{k}}^{\rm{H}},
\end{align}where ${\boldsymbol \Lambda}_{{\bf{X}}_k}$ is a rectangular diagonal matrix with the same rank as ${\boldsymbol \Lambda}_{{\boldsymbol A}_{k}}$ and $
1/b_k{\boldsymbol \Lambda}_{{\boldsymbol{\mathcal H}}_k}{\boldsymbol \Lambda}_{{\bf{X}}_k}={\boldsymbol \Lambda}_{{\boldsymbol A}_{k}}$. The scalar $b_k$ is chosen to make that ${\rm{Tr}}({\bf{\hat {F}}}_k{\bf{\hat { F}}}_k^{\rm{H}})=P_k$ holds.

Using $\textbf{Lemma 12}$ in \cite{Palomar03}, we can show that $
{\bf{\hat F}}_k^{\rm{H}}{\boldsymbol {\mathcal{H}}}_k^{\rm{H}}{\boldsymbol {\mathcal{H}}}_k{\bf{\hat F}}_k
\succeq {\bf{\tilde F}}_{k,{\rm{opt}}}^{\rm{H}}{\boldsymbol {\mathcal{H}}}_k^{\rm{H}}{\boldsymbol {\mathcal{H}}}_k
{\bf{\tilde F}}_{k,{\rm{opt}}}$. Together with the formulation of $\gamma_n(\{{\bf{\tilde F}}_k\}_{k=1}^K)$ in (\ref{OPT_F_2}), it can be concluded that $\gamma_n(\{{\bf{\hat F}}_k\}_{k=1}^K)\ge \gamma_n(\{{\bf{\tilde F}}_{k,{\rm{opt}}}\}_{k=1}^K)$. Since $g(\bullet)$ is an decreasing function, $g[{\boldsymbol \gamma}(\{{\bf{\hat F}}_k\}_{k=1}^K)]\le g [{\boldsymbol \gamma}(\{{\bf{\tilde F}}_{k,{\rm{opt}}}\}_{k=1}^K)]$.
Because ${\bf{\tilde F}}_{k,{\rm{opt}}}$ is the optimal solution, it is impossible to have $g[{\boldsymbol \gamma}(\{{\bf{\hat F}}_k\}_{k=1}^K)]< g [{\boldsymbol \gamma}(\{{\bf{\tilde F}}_{k,{\rm{opt}}}\}_{k=1}^K)]$. Therefore, ${\bf{\hat F}}_k$ must be the optimal solution. Furthermore, based on the relationship between of ${\bf{\tilde F}}_k$ and ${\bf{F}}_k$, it follows that
\begin{align}
\label{Apped_X_OPT}
{\bf{ F}}_{k,{\rm{opt}}}=\sqrt{\eta_{f_k}}(\alpha_k P_k{\boldsymbol
\Psi}_k+\sigma_{n_k}^2{\bf{I}}_{N_{T,k}})^{-{1}/{2}}{\bf{V}}_{{\boldsymbol {\mathcal{H}}}_k}{\boldsymbol \Lambda}_{{\bf{X}}_k}{\bf{V}}_{{\boldsymbol{A}}_k}^{\rm{H}}.
\end{align}Notice that in general the unitary matrix ${\bf{V}}_{{\boldsymbol{A}}_k}$ depends on the optimal solution ${\bf{\tilde F}}_{k,{\rm{opt}}}$. However, from (\ref{OPT_F_2}), it can be seen that the value of ${\bf{V}}_{{\boldsymbol{A}}_k}$
does not affect the objective functions and therefore it can be an arbitrary unitary matrix. Meanwhile, as the minimum dimension of ${\bf{\tilde F}}_{k}^{\rm{H}}{\boldsymbol {\mathcal{H}}}_k^{\rm{H}}{\boldsymbol {\mathcal{H}}}_k
{\bf{\tilde F}}_{k}$ is $N$, only $N\times N$ principal submatrix of ${\boldsymbol \Lambda}_{{\bf{X}}_k}$ can be nonzero. For notational convenience, we denote that $[{\boldsymbol \Lambda}_{{\bf{X}}_k}]_{1:N,1:N}={\boldsymbol \Lambda}_{{\boldsymbol {\mathcal F}}_k}$.

Substituting (\ref{Apped_X_OPT}) into the definition of $\eta_{f_k}$ in (\ref{eta_f}), we obtain a simple linear function of ${\eta}_{f_k}$, and $\eta_{f_k}$ can be easily solved to be
\begin{align}
\label{final} \eta_{f_k}&={\sigma_{n_k}^2}/\{1-\alpha_k{\rm{Tr}}[{\bf{V}}_{{\boldsymbol
{\mathcal{H}}}_k,N}^{\rm{H}}(\alpha_k P_k{\boldsymbol
\Psi}_k+\sigma_{n_k}^2{\bf{I}}_{N_{T,k}})^{-1/2} \nonumber \\
& \ \ \ \ \ \ \ \times {\boldsymbol \Psi}_k(\alpha_k P_k{\boldsymbol
\Psi}_k+\sigma_{n_k}^2{\bf{I}}_{N_{T,k}})^{-1/2}{\bf{V}}_{{\boldsymbol {\mathcal{H}}}_k,N}{\boldsymbol{\Lambda}}_{{\boldsymbol{\mathcal F}}_k}^2]\}\nonumber \\
&\triangleq {\xi}_k({\boldsymbol { \Lambda}}_{{\boldsymbol{\mathcal{F}}}_k}).
\end{align}

%


\begin{thebibliography}{99}



%


%

\bibitem{Swami2007}
A. Swami, Q. Zhao, Y.-W. Hong, and L. Tong, {\em Wireless Sensor Networks: Signal Processing and Communications}, Wiley Press, 2007.


\bibitem{Medina07}
O. Munoz-Medina, J. Vidal, and A. Agustin, ``Linear transceiver
design in nonregenerative relays with channel state information,''
{\em IEEE Trans. Signal Process.}, vol. 55, no. 6, pp. 2593--2604,
June  2007.


\bibitem{Tang07}
X. Tang and Y. Hua, ``Optimal design of non-regenerative MIMO
wireless relays,'' {\em IEEE Trans. Wireless Commun.}, vol. 6, no.
4, pp. 1398--1407, Apr. 2007.

\bibitem{Guan08}
W. Guan and H. Luo, ``Joint MMSE transceiver design in
non-regenerative MIMO relay systems,'' {\em IEEE Commun.
Lett.}, vol. 12, no. 7, pp. 517--519, July 2008.

\bibitem{Tseng09}
F.-S. Tseng, W.-R. Wu, and J.-Y. Wu ``Joint source/relay precoder design in nonregenerative cooperative systems using an MMSE criterion,'' {\em IEEE Trans. Wireless Commun.}, vol. 8, no. 10, pp. 4928--4933, Oct. 2009.

\bibitem{Mo09}
R. Mo and Y. Chew,
``Precoder design for non-regenerative MIMO relay systems,'' {\em IEEE Trans. Wireless Commun.}, vol. 8, no.
10, pp. 5041--5049, Oct. 2009.


\bibitem{Xing10}
C. Xing, S. Ma, and Y.-C. Wu, ``Robust joint design of linear relay precoder and destination equalizer for dual-hop amplify-and-forward MIMO relay systems,'' {\em IEEE
Trans. Signal Process.}, vol. 58, no. 4, pp. 2273--2283, Apr. 2010.

\bibitem{Xing1012}
C. Xing, S. Ma, Y.-C. Wu, and T.-S. Ng, ``Transceiver design for dual-hop nonregenerative MIMO-OFDM relay systems under channel uncertainties,'' {\em IEEE Trans. Signal Process.}, vol. 58, no. 12, pp. 6325--6339, Dec. 2010.


\bibitem{Rong09}
Y. Rong, X. Tang, and Y. Hua, ``A unified framework for optimizing
linear nonregenerative multicarrier MIMO relay communication
systems,'' {\em IEEE Trans. Signal Process.}, vol. 57, no. 12, pp.
4837--4851, Dec. 2009.



\bibitem{XingICSPCC}
C. Xing, Z. Fei, Y.-C. Wu, S. Ma, and J. Kuang, ``Robust transceiver design for AF MIMO relaying systems with column correlations,'' {\em in Proc. IEEE ICSPCC 2011}, Xi'an, China, Sep. 2011.


\bibitem{XingGlobecom}
C. Xing, S. Ma, Z. Fei, Y.-C. Wu, and J. Kuang, ``Joint robust weighted LMMSE transceiver design for dual-hop AF multiple-antenna relay systems,'' {\em in Proc. IEEE Globecom 2011}, Houston, TX, USA, Dec. 2011.


\bibitem{Rong2009TWC}
Y. Rong and Y. Hua, ``Optimality of diagonalization of multi-hop MIMO relays,'' {\em IEEE Trans. Wireless Commun.}, vol. 8, no. 12,
pp. 6068--6077, Dec. 2009.





\bibitem{Jiang2005}
Y. Jiang, J. Li, and W. W. Hager, ``Uniform channel decomposition for MIMO communications,'' {\em IEEE Trans. Signal Process.}, vol. 53, no. 11, pp. 4283--4294, Nov. 2005.


\bibitem{Shenouda2008}
M. B. Shenouda and T. N. Davidson,``A framework for designing MIMO systems with decision feedback equalization or Tomlinson-Harashima precoding,''
{\em IEEE J.  Select. Areas Commun.}, vol. 26 no. 2 pp.401--411, Feb. 2008.


\bibitem{Amico2008}
A. A. D'Amico, ``Tomlinson-Harashima precoding in MIMO systems: A unified approach to transceiver optimization based on multiplicative Schur-convexity,'' {\em IEEE Trans. Signal Process.}, vol. 56, no. 8, pp. 3662--3677, Aug. 2008.


\bibitem{Rong2011}
Y. Rong and M. R. A. Khandaken, ``On Uplink-downlink duality of multi-hop MIMO relay channel,'' {\em IEEE Trans. Wireless Commun.}, vol. 10, no. 6, pp. 1923--1931, June 2011.

\bibitem{Palomar2006}
D. P. Palomar and Y. Jiang, ``MIMO Transceiver Design via Majorization Theory,'' {\em Foundations and Trends in Communications and Information Theory}, Now Publishers, vol. 3, no. 4, pp. 331--551, Nov. 2006.


\bibitem{Dietrich2007}
F. A. Dietrich, P. Breun and W. Utschick,  ``Robust Tomlinson-Harashima precoding for the wireless broadcast channel,'' {\em IEEE Trans. Signal Process.}, vol. 55, no. 2, pp. 631--644, Feb. 2007.


\bibitem{Fischer2002}
R. F. H. Fischer,  {\em Precoding and Signal Shaping for Digital Transmission}. New York: Wiley-IEEE, July 2002.


\bibitem{Zhang08}
X. Zhang, D. P. Palomar, and B. Ottersten, ``Statistically robust
design of linear MIMO transceivers,'' {\em IEEE Trans. Signal
Process.}, vol. 56, no. 8, pp. 3678--3689, Aug. 2008.



\bibitem{Ding09}
M. Ding and S. D. Blostein, ``MIMO minimum total MSE transceiver
design with imperfect CSI at both ends,'' {\em IEEE Trans. Signal
Process.}, vol. 57, no. 3, pp. 1141--1150, Mar. 2009.

\bibitem{Ding10}
M. Ding and S. D. Blostein, ``Maximum mutual information design for MIMO systems with imperfect channel knowledge,'' {\em IEEE Trans. Inf. Theory}, vol. 56, no. 10, pp.4793--4801, Oct. 2010.


\bibitem{Palomar03}
D. P. Palomar, J. M. Cioffi, and M. A. Lagunas, ``Joint Tx-Rx beamforming design for multicarrier MIMO channels: A unified framework for convex optimization,'' {\em IEEE Trans. Signal
Process.}, vol. 51, no. 9, pp. 2381--2401, Sep. 2003.
















%

%

%





\bibitem{Marshall79}
A. W. Marshall and I. Olkin, {\em Inequalities: Theory of
Majorization and Its Applications}. New York: Academic Press, 1979.








\bibitem{Weyl1949}
H. Weyl, ``Inequality between the two kinds of eigenvalues of a linear transformation,'' {\em Proc. Nat. Acad. Sci. USA}, vol. 35, no. 7, pp. 408--411, July 1949.



















\bibitem{XingWCSP}
C. Xing, Z. Fei, S. Ma, J. Kuang, and Y.-C. Wu, ``Robust linear transceiver design for multi-hop non-regenerative MIMO relaying systems,'' {\em in Proc. IEEE WCSP 2011}, Nanjing, China, Nov. 2011.

\bibitem{Boyd04}
S. Boyd and L. Vandenberghe, {\em Convex Optimization}. Cambridge
University Press, 2004.


\bibitem{Yu04}
W. Yu, W. Rhee, S. Boyd, and J. Cioffi, ``Iterative water-filling for Gaussian vector multiple access channels,'' {\em IEEE Trans. Infor. Theory}, vol. 50, no. 1, pp.145--152, Jan. 2004.


\bibitem{Zhang2011}
W. Zhang, U. Mitra, and M. Chiang, ``Optimization of amplify-and-forward multicarrier two-hop transmission,'' {\em IEEE Trans. Commun.}, vol. 59, no. 5, pp. 1434--1445, May 2011


\bibitem{Musavian07}
L. Musavian, M. R. Nakhi, M. Dohler, and A. H. Aghvami, ``Effect of
channel uncertainty on the mutual information of MIMO fading
channels,'' {\em IEEE Trans. Veh. Technol.}, vol. 56, no. 5, pp.
2798--2806, Sep. 2007.




\bibitem{Yoo2005}
T. Yoo and A. Goldsmith, ``Capacity and power allocation for fading MIMO channels with channel estimation error,'' {\em IEEE Trans. Inf. Theory  } vol. 52, no. 5, pp. 2203--2214, May 2006.



\bibitem{Zhang2005}
J.-K. Zhang, A. Kav$\check{\rm{c}}$i$\acute{\rm{c}}$, and K. M. Wong,  ``Equal-diagonal QR decomposition and its application to precoder design for successive-cancellation detection,'' {\em IEEE Trans. Inf. Theory}, vol. 51, no. 1, pp. 154--172, Jan. 2005.

\bibitem{Zhuhuiling2009}
H. Zhu and J. Wang, ``Chunk-Based Resource Allocation in OFDMA Systems-Part I Chunk Allocation,'' {\em IEEE Trans. Commun.}, vol. 57, no. 9, pp. 2734--2744, Sep. 2009.

\bibitem{zhouyiqing2008}
Y. Zhou, J. Wang, T.S. Ng, K. Higuchi, and M. Sawahashi, ``OFCDM: a promising broadband wireless access technique,'' {\em IEEE Commun. Magazine}, vol. 46, no.3, pp. 39--49, March 2008.

\bibitem{zhouyiqing2005}
Y. Zhou, J. Wang, and M. Sawahashi, ``Downlink transmission of broadband OFCDM systems--Part I: Hybrid Detection,'' {\em IEEE Trans. Commun.}, vol. 53, no. 4, pp. 718-729, April 2005.


\end{thebibliography}
\end{document}